\documentclass[prl,twocolumn,superscriptaddress,floatfix,noshowpacs,10pt,longbibliography]{revtex4-2}
\usepackage{graphicx,bm,times}
\usepackage{amsmath}
\usepackage{amsfonts}
\usepackage{amssymb}
\usepackage{bm}
\usepackage{color}
\usepackage{times}
\usepackage{xcolor}
\usepackage[%
  colorlinks=true,
  urlcolor=blue,
  linkcolor=blue,
  citecolor=blue
  ]{hyperref}

\begin{document}

\title{Unconventional localization of electrons inside of a  nematic electronic phase }

\author{L. Farrar}
\affiliation{Centre for Nanoscience and Nanotechnology, Department of Physics, University of Bath, Bath, BA2 7AY, United Kingdom }

\author{Z. Zajicek}
\affiliation{Clarendon Laboratory, Department of Physics,
	University of Oxford, Parks Road, Oxford OX1 3PU, UK}

\author{A.B. Morfoot}
\affiliation{Clarendon Laboratory, Department of Physics,
	University of Oxford, Parks Road, Oxford OX1 3PU, UK}

\author{M. Bristow}
\affiliation{Clarendon Laboratory, Department of Physics,
University of Oxford, Parks Road, Oxford OX1 3PU, UK}

\author{O. S. Humphries}
\affiliation{Clarendon Laboratory, Department of Physics,
University of Oxford, Parks Road, Oxford OX1 3PU, UK}

\author{A. A.\;Haghighirad}
\affiliation{Clarendon Laboratory, Department of Physics,
University of Oxford, Parks Road, Oxford OX1 3PU, UK}
\affiliation{Institute for Quantum Materials and Technologies, Karlsruhe Institute of Technology, 76021 Karlsruhe, Germany}

\author{A. McCollam}
\affiliation{High Field Magnet Laboratory (HFML-EMFL), Radboud University, 6525 ED Nijmegen, The Netherlands}

\author{S. J. Bending}
\affiliation{Centre for Nanoscience and Nanotechnology, Department of Physics, University of Bath, Bath, BA2 7AY, United Kingdom }

\author{A. I. Coldea}
\email[corresponding author:]{amalia.coldea@physics.ox.ac.uk}
\affiliation{Clarendon Laboratory, Department of Physics,
University of Oxford, Parks Road, Oxford OX1 3PU, UK}

\date{\today}

\begin{abstract}
The magnetotransport behaviour inside the nematic phase of bulk FeSe
reveals unusual multiband effects that cannot be reconciled
with a simple two-band approximation proposed by surface-sensitive spectroscopic probes.
In order to understand the role played by the multiband electronic structure and the degree of two-dimensionality
we have investigated the electronic properties of exfoliated flakes of FeSe by reducing their thickness.
Based on magnetotransport and Hall resistivity measurements, we assess the mobility spectrum
 that suggests an unusual asymmetry between the mobilities of the electrons and holes
 with the electron carriers becoming  localized inside the nematic phase.
 Quantum oscillations  in magnetic fields up to 38~T indicate the presence of a hole-like quasiparticle with
a lighter effective mass and a quantum scattering time three times shorter, as compared with bulk FeSe.
  The observed localization of negative charge carriers by
  reducing dimensionality can be driven by orbitally-dependent correlation effects, enhanced interband spin-fluctuations
or a Lifshitz-like transition which affect mainly the electron bands.
The electronic localization leads to a fragile two-dimensional superconductivity in thin flakes of  FeSe,
in contrast to the two-dimensional high-$T_{\rm c}$ induced with electron doping via dosing or using a suitable interface.
\end{abstract}

\maketitle

Among different classes of unconventional superconductors,
iron-based systems display rich physics due to their
multiband structure and the competition between
different  electronic interactions \cite{Fernandes2022}.
Iron-chalcogenides are among the most
strongly correlated  iron-based superconductors and,
due to the large intra-atomic exchange caused by the Hund's coupling,
 the correlation strengths are expected
to be strongly orbitally dependent \cite{Yin2011}.
This orbital differentiation can lead to an orbital-selective Mott transition,
 or spectral weight transfer, where the band with dominant
$d_{xy}$ orbital appears more  insulating while other bands
with $d_{xz}$ and  $d_{yz}$ orbital character  remain metallic \cite{Fernandes2022,Cai2020}.
As a result,  electronic and superconducting properties are likely to be influenced
by these effects, as seen in orbitally-dependent band shifts in angle-resolved
photoemission spectroscopy (ARPES) in iron-chalogenides \cite{Yi2015corr,ZKLiu2015}
and orbital-dependent Cooper pairing  \cite{Sprau2017}.

FeSe is a candidate system in which the presence of the lower-Hubbard band
establishes the important role of electronic correlations, orbital-dependent
band shifts in the nematic phase  and Fermi surface shrinking
\cite{Watson2015a,Fanfarillo2016,Watson2017a,Coldea2017}.
The strength of these effects can be suppressed by isoelectronic substitution with sulphur
\cite{Watson2015c,Reiss2017,Coldea2021}.
The interatomic Coulomb repulsion in FeSe can  produce
 a strongly renormalized low-energy band structure
where the van Hove singularity sits remarkably close to the Fermi level
in the high-temperature electron liquid phase \cite{Jiang2016}.
 ARPES studies under strain
suggest that in FeSe one electron pocket either is
missing or the spectral weight is transferred between its two electron pockets
\cite{Watson2017,Cai2020}.
Furthermore, neutron scattering suggests the
coexistence of both stripe and
N\`{e}el spin fluctuations with a substantial amount of spectral weight
transferred towards stripe spin fluctuations inside the nematic phase \cite{Wang2016}.
The lack of long-range antiferromagnetic order in FeSe has been linked
to the competition between different types of magnetic order
that can lead to significant magnetic frustration \cite{Glasbrenner2015}.

Transport properties of systems with orbitally-dependent correlations are predicted
to display a coherence-incoherence crossover as a function of temperature \cite{Haule2009}
and, in addition, nematic iron-based superconductors  are prone to anisotropic single-particle
scattering enhanced by interband spin or charge fluctuations \cite{Breitkreiz2013}.
In the presence of spin fluctuations, the scattering
is strongly influenced by quasiparticles close to {\it hot spots}
at Fermi surface locations where the nesting is strong
\cite{Rosch2000,Koshelev2016}.
Furthermore, the large fluctuating moments
with N\`{e}el and stripe magnetic instabilities \cite{Wang2016}
are likely to strongly affect the magnetotransport behaviour.
 Magnetotransport studies of FeSe  and FeSe$_{1-x}$S$_x$
 have identified  a linear resistivity regime  and a large magnetoresistance
inside the nematic phase \cite{Bristow2020,Watson2015b}.
 However, the low-field magnetotransport data in the normal state suggest that,
in addition to one hole and one almost compensated electron band,
the nematic phase of FeSe exhibits an additional tiny electron pocket
with a high mobility and non-linear Hall coefficient  \cite{Huynh2014,Watson2015b}
that can also be found in the low pressure regime \cite{Terashima2016}.
Additionally, the amplitude of the quantum oscillations at low temperatures and
high magnetic fields indicates that hole carriers are likely to be more mobile than electron
bands \cite{Watson2015b}, and the highly mobile holes and enhanced spin fluctuations
also dominate the high-$T_{\rm c}$ pressure phase of FeSe  \cite{Sun2016pressure}.

{\color{blue} Among iron-based superconductors, FeSe
displays an anomalous electronic nematic state, strong
electronic correlations and orbitally-dependent
band shifts that can influence its superconducting pairing.
Here, we report a detailed magnetotransport study
of thin flakes of FeSe
that reveals unconventional transport
in which the hole carriers remain highly mobile,
whereas the mobility of the electron carriers is low, with hardly any temperature dependence, inside the nematic phase.
This suggests an unusual localization of negative charge carriers
that may be caused by orbital-dependent
enhanced correlations, scattering of spin fluctuations and/or a topological electronic transition.
 As the superconductivity is suppressed by reducing the flake thickness,
it suggests that the electron pockets participate actively in the superconducting pairing.
By doping, electron pockets expand enabling the high-Tc superconductivity.}

In this paper, we present a detailed magnetotransport study of exfoliated thin flakes of FeSe as compared with  bulk single crystals.
The mobility spectrum and a two-band model reveal that the mobility of the holes in thin flake devices
 is much higher than that of electrons.
Electrons become localized at low temperatures, due to the enhancement of the
 orbitally-dependent correlations and enhanced anisotropic scattering in two-dimensional devices.
 From quantum oscillations  we find that the size of the extremal hole orbit is
 smaller and the effective mass of the hole band is lighter than in bulk.
In the low-temperature regime,
the resistivity shows a linear dependence down
to the lowest temperatures but Fermi liquid
behaviour is restored in the cleanest flake, where quantum oscillations are present.
While the thinner devices  could be sensitive to increased impurity and surface scattering,
the reduction in $T_{\rm c}$ in thin flakes is directly correlated
with the suppression of the nematic phase and the localization of electron carriers.

\begin{figure}[t!]
\centering
\includegraphics[trim={0cm 0cm 0cm 0cm}, width=1\linewidth,clip=true]{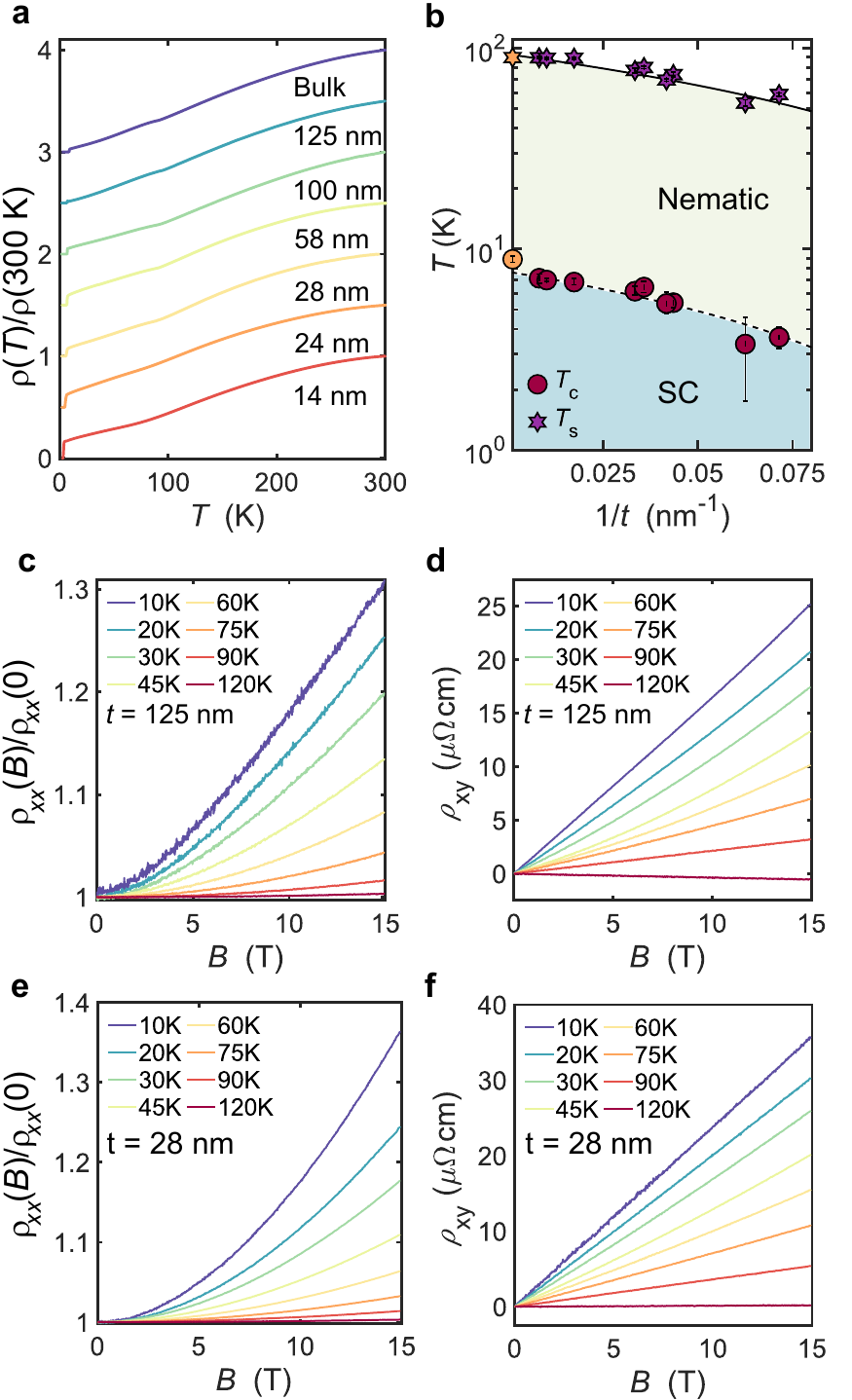}
\caption{{ Magnetotransport of thin flakes of FeSe.}
(a) Temperature dependence of the normalised zero-field resistance $\rho$($T$)/$\rho$(300K) for
a bulk crystal and different thin flake devices ($t$ = 14-125~nm). Data are vertically offset and this panel is updated with new data
from Ref.~\cite{Farrar2020}.
 (b) Inverse thickness $t$ dependence of the superconducting transition temperature, $T_{\mathrm{c}}$, and the structural transition temperature, $T_{\mathrm{s}}$.
 Bulk values are shown on the left with light orange symbols. The dashed line is a fit to the Cooper model  and the solid line is a guide to the eye
  for the changes in $T_{\rm s}$ with the thickness $t$ \cite{Cooper1961,Simonin1986}.
 (c-f) The magnetic field dependence of the longitudinal
  and transverse components of the resistivity, $\rho_{xx}$ and $\rho_{xy}$,
  at different constant temperatures
 for two devices with $t$=125 and $t$ = 28~nm.}
 \label{fig:Figure_1}
\end{figure}

\section*{Results}

\subsection*{Transport properties of thin flakes of FeSe. }
Fig.~\ref{fig:Figure_1}(a) shows the temperature dependence of the normalized zero-field resistance,
$\rho$($T$)/$\rho$(300K), for a bulk crystal and six different thin flake
 devices with thicknesses in the range $t=$14-125~nm.
Bulk FeSe undergoes a tetragonal to orthorhombic distortion
at $T_{\rm s} \approx 90$~K without any accompanying
long-range magnetic order, followed by the onset of superconductivity at  $T_{\rm c} \sim 9$~K \cite{Bohmer2013}.
These parameters in single crystals are sensitive to the growth conditions and the impurity level;
the suppression of  $T_{\rm c}$ is affected by the increase in the amount of  disorder,
 as measured by the residual resistivity ratio ($RRR$),
which linearly correlates with the suppression of $T_{\rm s}$ \cite{Bohmer2016}.
In thin flakes, the superconducting transition temperature, $T_{\rm c}$, is already lowered
from the bulk single crystal value to 7.2~K for a $t=125$~nm device, decreasing
 further to 3.6~K for a $t=14$~nm device, as reported previously \cite{Farrar2020}.
The suppression of superconductivity in the $t=125$~nm device occurs despite the high residual resistance ratio
of $RRR \sim 32 $, which is similar to  bulk crystals from the same batch \cite{Bristow2020}.
In thinner flakes, the $RRR$ value falls as a function of decreasing thickness, reducing to 5.5 in the $t=14$~nm device
as shown in Fig.~S4(c) in the SI Appendix \cite{SM},
being similar to the effect of impurity scattering by Cu doping in FeSe \cite{Gong2021,Zajicek2021Cu}.
Similarly, the emergence of the nematic phase at $T_{\mathrm{s}}$, which results in significant in-plane distortion of the Fermi surface and
orbitally-dependent band shifts \cite{Coldea2017}, becomes smeared and less defined for thinner flakes.
 Fig.~\ref{fig:Figure_1}(b) shows the variation of $T_{\mathrm{s}}$ and
 $T_{\mathrm{c}}$ as a function of inverse thickness ($1/t$) for different devices,
  revealing that both are suppressed for thinner flakes; interestingly,  we find a linear dependence
  between $T_{\mathrm{s}}$ and $T_{\mathrm{c}}$, as found in Cu-substituted FeSe \cite{Zajicek2021Cu} (see Fig.~S4(b)  in SI Appendix \cite{SM}).
    As these two parameters are correlated  and are suppressed as
    the $RRR$ ratio is reduced, it suggests that  the two-dimensional confinement, enhanced fluctuations and
   surface impurity scattering,
   play an important role in the suppression of $T_{\rm c}$ in thin flakes of FeSe \cite{Farrar2020}.

\subsection*{Low-field magnetotransport behaviour}
Figs.~\ref{fig:Figure_1}(c-f) show the field dependence of the longitudinal magnetoresistance normalized by the
zero-field value, $\rho_{xx}$($B$)/$\rho_{xx}(0)$,
and the Hall resistivity, $\rho_{xy}$, for two different devices with $t=$125 and 28~nm, respectively.
FeSe is a multi-band stoichiometric compound in which charge compensation requires
 that $n = n_{e} = n_{h}$.
At high temperatures in the tetragonal phase,
 the magnetotransport properties of bulk FeSe can be accurately described using
 a compensated two-band model,
  as  detailed in the SI Appendix \cite{SM}.
 This model corresponds to a hole pocket at the  Brillouin zone centre and an electron pocket at the
corner of the Brillouin zone \cite{Watson2015b}.
 This two-band picture also describes the high temperature magnetotransport behaviour of thin flakes
 whereby the magnetoresistance does not saturate and  $\rho_{xy}$  shows a linear
 dependence on magnetic field (see Fig.~S5 in the SI Appendix \cite{SM}).

\begin{figure}[t!]
\centering
\includegraphics[trim={0cm 0cm 0cm 0cm}, width=1\linewidth,clip=true]{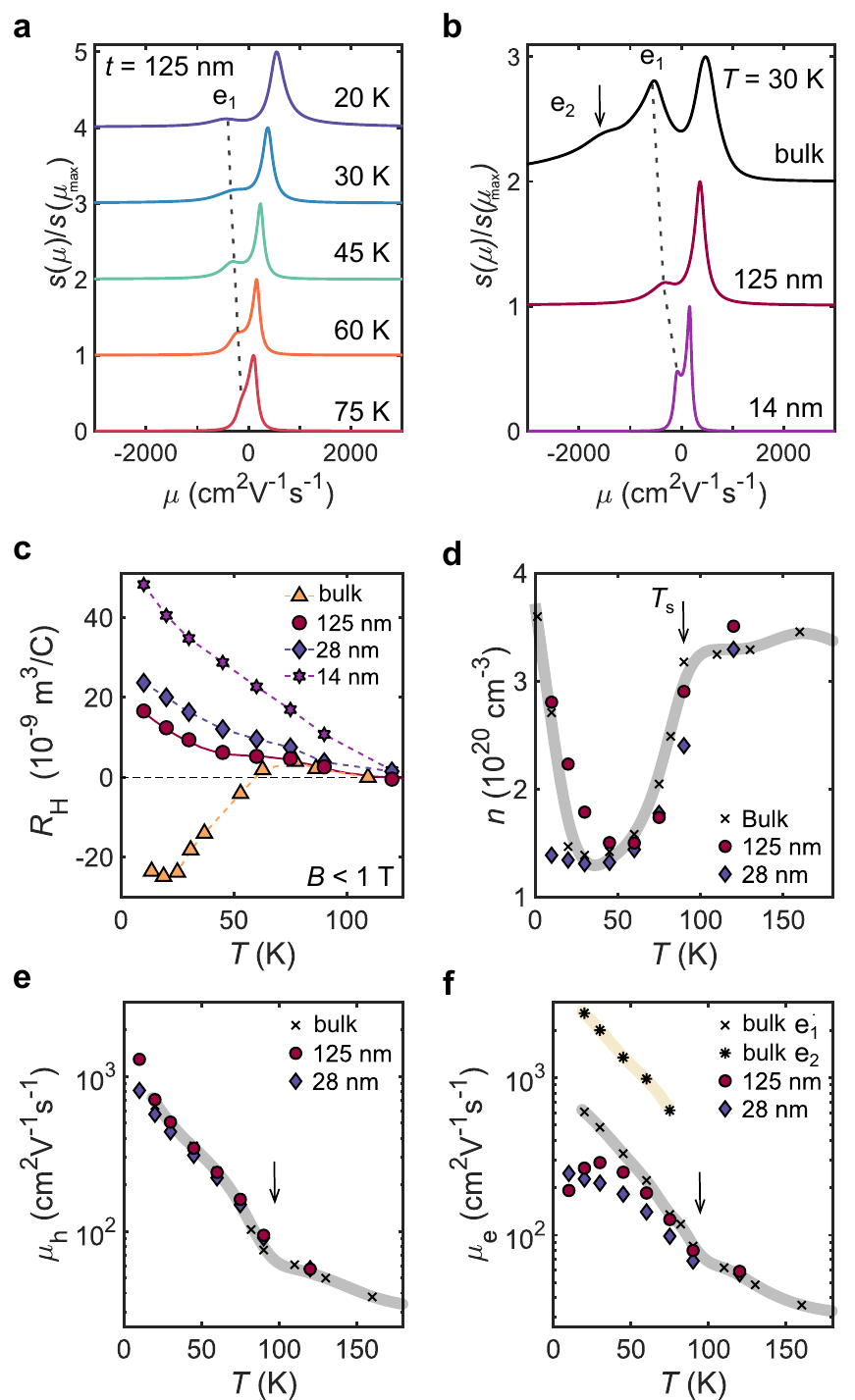}
\caption{{The changes in the carrier mobilities with temperature.}
 (a) Temperature dependence of the normalized mobility spectra to the maximum value $s(\mu_{max})$
  of electron-like (negative) and hole-like (positive) carriers for a $t=$125~nm device.
 (b)  A comparison of the normalized mobility spectra to the to the maximum value $s(\mu_{max})$
  for a bulk crystal and different thin flake devices of FeSe at $T = 30$~K.
 The spectra are vertically offset for clarity.
 Dotted line shows the mobilities of the electron  charge carrier $e_1$ and the arrow indicates the highly mobile electron, $e_2$, found only in the bulk.
 (c) The temperature dependence of the low field Hall coefficient $R_{\rm H}$ (extracted for $B<$ 1~T) for a bulk FeSe crystal and three thin flake devices.
 d) Temperature dependence of the carrier density   extracted using a compensated two-carrier model $n=n_e=n_h$ (see SI Appendix \cite{SM}).
  Temperature dependence of carrier mobilities
  for (e)  the hole-like carriers, $\mu_h$, and (f) electron-like carriers,  $\mu_e$,  for various samples.
  The data points for the bulk single crystals of FeSe and
   the three-carrier model parameters are from Ref.~\cite{Watson2015b} and the solid lines are guides to the eye.
   The vertical arrows indicate the position of $T_{\rm s}$.
   }
 \label{fig:Figure_2}
\end{figure}

Inside the nematic  phase (below 75~K), the Hall component $\rho_{xy}$  of
 bulk crystals of FeSe displays non-linear behaviour in magnetic field
and a negative slope  \cite{Watson2015b}
  and a deviation from the compensated two-band behaviour,
whereas all  thin flake devices display a positive Hall component in magnetic field
(see also Fig.~S5  in the SI Appendix \cite{SM}).
To visualize the differences between the bulk and thin flakes,
 Fig.~\ref{fig:Figure_2}(c) shows  the temperature dependence of the
  Hall coefficient, $R_{\mathrm{H}}=\rho_{xy}/B$ in low magnetic fields ($B <$ 1T).
All the thin flake devices have a positive Hall coefficient below $T_{\mathrm{s}}$
suggesting that the transport behaviour becomes increasingly
 dominated by the hole-like carriers.
We detect a local maximum in $R_{\rm H}$ around 65~K for the $t=125$~nm,
which is close to the temperature at which the Hall coefficient of bulk FeSe \cite{Watson2015b}
and also the resistivity anisotropy under strain changes sign \cite{Ghini2021}.
These striking changes in transport behaviour could signify
the development of anisotropic scattering effects  inside the nematic phase
which can be enhanced
by lowering the temperature and reducing the thickness.
The $\rho_{xy}(B)$ component
deviates from a linear dependence in magnetic field in the same temperature
regime (see corresponding derivatives in Fig.~S5 in the SI Appendix \cite{SM}).
However, linear dependence is detected
at the lowest  temperatures,
in the regime of isotropic scattering caused mainly by impurities,
and at the highest temperatures, where electron-phonon scattering becomes important,
similar  to other reports on thin flakes \cite{Lei2016}.
    Overall, the behaviour of the Hall coefficient in thin flakes of FeSe is in stark contrast to bulk FeSe,
  in which $R_{\mathrm{H}}$ becomes negative below the nematic transition, as seen in Fig.~\ref{fig:Figure_2}c.

\subsection*{Mobility spectrum}
To analyse in detail the origin of the changes inside the nematic phase,  we perform a mobility-spectrum analysis of the magnetotransport data (Fig.~\ref{fig:Figure_2}).
This approach has previously been successful in characterising the transport behaviour of bulk FeSe \cite{Huynh2014}  and other
multi-band systems \cite{Zhao2021}.  The modelling is based on the methodology detailed in Ref.~\cite{Beck1987,Humphries2016},
 which eliminates the need for making {\it a priori} assumptions on the transport parameters in a multicarrier system
 and describes the mobility spectrum,  $s$($\mu$)=$e \mu n(\mu)$,
 with the negative electron charge assigned to a negative value of mobility.
 Fig.~\ref{fig:Figure_2}(a) shows the evolution of the
 normalised mobility spectrum $s$($\mu$)/$s$($\mu_{max}$)  for the $t=125$~nm device as a function of temperature.
 The height of $s$($\mu$) could be linked to the variation of the
 number of carriers, $n(\mu)$, whereas the width of the peak of $s(\mu)$
 relates to a distribution of relaxation times and
 the magnetic field resolution, such that $\mu B <1 $ \cite{Beck1987}
 (see Fig.~S9  in the SI Appendix \cite{SM}).
 As a function of temperature, the mobility of
 positive charge carriers  increases strongly with decreasing temperature reaching $\mu_{h} \sim1200$~cm$^{2}$V$^{-1}$s$^{-1}$ at 10~K.
 On the other hand, the electron mobility has a much weaker increase with decreasing temperature, and below 50~K
 the corresponding mobility peak position
   only shifts slightly from $|\mu_{e1} |\sim300(50)$~cm$^{2}$V$^{-1}$s$^{-1}$.
  As the mobility of the holes is larger than that of electrons,
 it explains the positive sign of the Hall coefficient in thin flakes,
 shown in Fig.~\ref{fig:Figure_2}(c).

 Next, we compare the mobility spectrum of two different flake devices with those
of the bulk crystal at 30~K, as shown in Fig.~\ref{fig:Figure_2}(b).
The bulk mobility spectrum has a complex shape for the electron-like charge carriers, that could indicate
the presence of an additional highly mobile electron band, $e_{2}$,
(or the existence of sharp changes in curvature of the Fermi surface for electron-like
flower shape pockets in Fig.~\ref{fig:Figure_4}(b)),
besides the mobility shoulder corresponding to the high carrier density electron band, $e_{1}$.
This behaviour is in agreement with a previous report indicating a  broad mobility spectrum for electrons,
with the highly mobile carriers assigned as ultra-fast Dirac-like carriers in bulk FeSe \cite{Huynh2014}.
This highly mobile carrier, $e_2$, is not visible in any of the thin flakes, even for the 125~nm flake
which has a similar value of $RRR$ to that of the bulk.
This suggests a high sensitivity of the electronic structure,  in particular the electron pockets, to reduced
interlayer coupling and enhanced two-dimensional confinement.
Fig.~\ref{fig:Figure_2}(b) also shows that the mobilities of both electron and hole carriers
 in general decrease with the thickness of the flakes
 and reduced RRR ratio (see Fig.~S4(c) in SI Appendix \cite{SM}).
 In the case of the thinnest flake with $t=14$~nm, both the  electron $e_{1}$ and
 hole mobilities have been drastically reduced, as compared to the bulk crystal
($|\mu_{e1}|$ = 80 and $\mu_{h}$ = 160 cm$^{2}$V$^{-1}$s$^{-1}$ at 30~K).

 To provide quantitative insights into the behaviour of the charge carriers in the thin flakes,  we simultaneously fit the two resistivity components, $\rho_{xx}$ and $\rho_{xy}$,
 to the compensated two-band model using the values from the mobility spectrum as starting parameters.
  Figs.~\ref{fig:Figure_2}(d-f) show the temperature dependence
  of the carrier density $n = n_{h} = n_{e}$, and the field-independent mobilities $\mu_h$ and $\mu_e$, compared with those for the bulk single crystals of FeSe \cite{Watson2015b}.
 At high temperatures, the extracted values for $n$ are similar to those of bulk samples, which show a relatively constant carrier density of $n \approx 3 - 4 \times 10^{20}$ cm$^{-3}$.
 Unexpectedly, the carrier density shows a marked reduction of more than a factor of 2
 at 45~K, before rapidly increasing back to the high temperature value below 10~K. This drop in carrier density below $T_{\rm s}$
 cannot be reconciled with the expectation that the Fermi surface pockets should only deform but not change size inside the nematic phase, in the absence of any spin-density wave order.
 This anomalous behaviour may be caused by the onset of strongly anisotropic scattering at the Fermi surfaces below $T_{\mathrm{s}}$ \cite{Watson2015c},
 arising from the presence of strong spin fluctuations \cite{Wiecki2018},
  as different parts of the Fermi surfaces that are nested by the antiferromagnetic ordering vector will experience
  a dramatic increase in the scattering rate \cite{Breitkreiz2014}.
The calculated drop in the effective carrier density, $n$,
  in the absence of any change in the Fermi surface volume,  could be a manifestation of strongly anisotropic scattering
  inside  the nematic phase or that some of the charge carriers become more localized
  and do not contribute to transport behaviour.
  However, $n$ recovers its value in the low temperature limit for $t$=125~nm flake where the
 isotropic impurity scattering dominates, similar to the bulk behaviour
  \cite{Watson2015b}.

Figs.~\ref{fig:Figure_2}(e,f) show the extracted mobilities from the two-band analysis which confirms the striking difference in the mobility behaviour
of the two types of charge carriers ($\mu_{h}$ and $\mu_{e1}$).
For all the measured thin flake devices,  the hole mobilities are much larger than those of electrons, exhibiting a similar temperature dependence to that of the bulk
   band. There is a slightly decreased mobility of $\mu_{h}$ = 810 cm$^{2}$V$^{-1}$ s$^{-1}$
   at $T= $10 K for  the $t=$28~nm sample,  attributed to the increasing importance of surface
   scattering due to the increasing
   surface-to-volume ratio in thinner samples
    as well as the reduction of the residual resistivity ratio  due to other extrinsic effects
    (additional scattering from charged centres in the SiO$_2$ substrate).
     In contrast to the behaviour of the holes, the electron mobility of thin flakes displays a much weaker temperature dependence
 that deviates significantly from the bulk $e_{1}$ value. Actually, the electron mobility plateaus in thin flake samples at low temperatures,
 as is clearly seen in the mobility data  of Fig.~\ref{fig:Figure_2}(f).
 The contrasting behaviour of the mobilities of the holes and electrons is unexpected and correlates with the suppression of superconductivity in thin flakes in two-dimensions.
 This behaviour in FeSe is different from that found in iron-pnictides where electrons  are often more mobile than holes \cite{Kasahara2012}.
 Our findings are in broad agreement with the results of terahertz spectroscopy in FeSe thin films
 that detect that the scattering time of the hole carrier becomes substantially longer than that of the electron at lower temperatures \cite{Yoshikawa2019}.
 Highly mobile hole carriers were also found in the high-$T_{\rm c}$ phase of bulk FeSe under pressure \cite{Sun2017}.

\begin{figure*}[t!]
\centering
\includegraphics[trim={0cm 0cm 0cm 0cm},width=1\linewidth,clip=true]{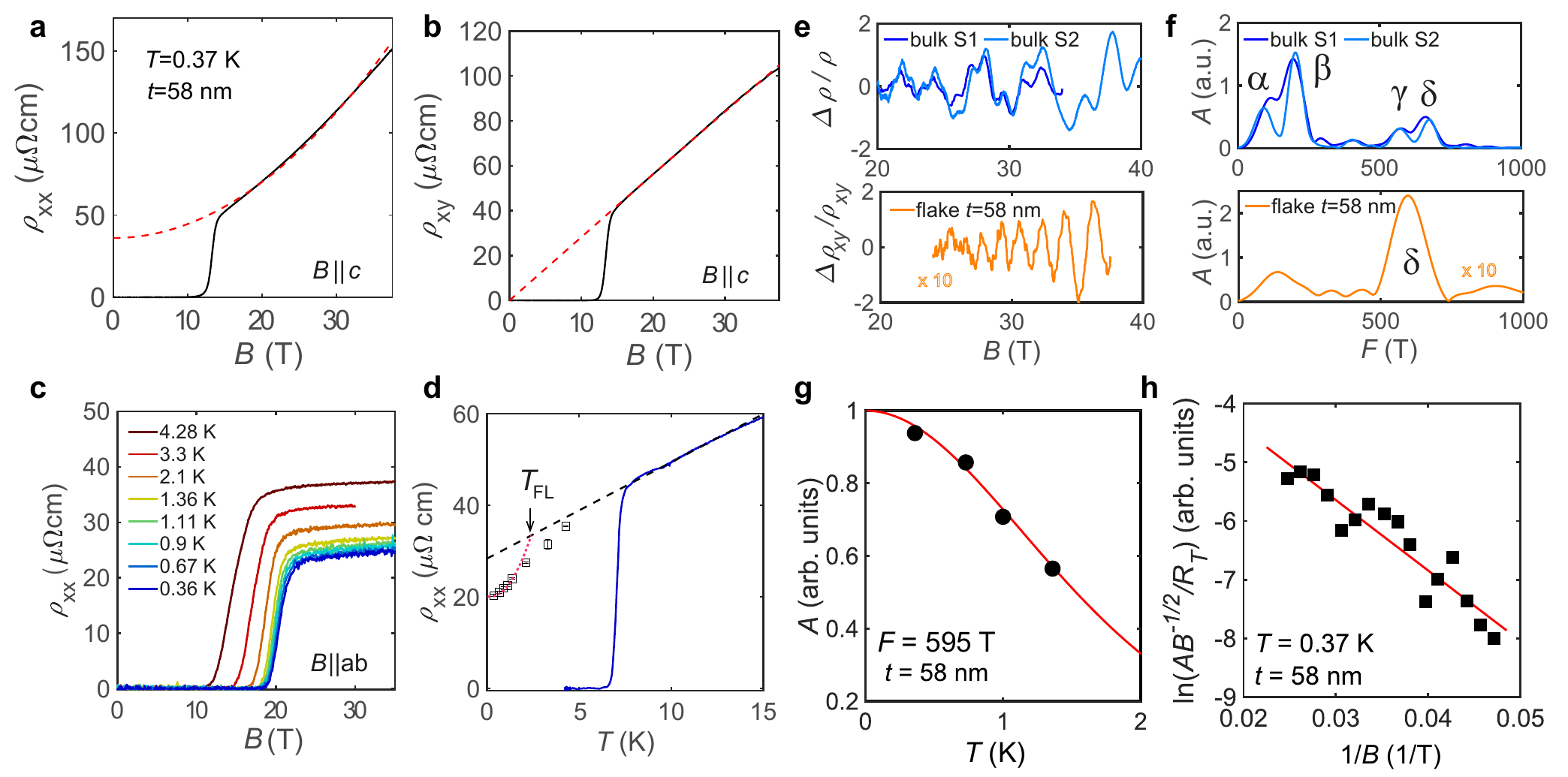}
\caption{{ High-field magnetotransport and quantum oscillations in a thin flake device of FeSe.}
  (a) Longitudinal resistivity  and (b) Hall resistivity as a function of $B$
  for a $t$ = 58 nm flake at $T$ = 0.37 K.
  The dashed lines are fits to a two-band model with
 $\mu_h$= 1017 cm$^2$V$^{-1}$s$^{-1}$ and $\mu_e$= 231 cm$^2$V$^{-1}$s$^{-1}$.
 (c) In-plane resistivity as a function of magnetic field at different fixed temperatures.
 (d) Low temperature zero-field resistivity together with extrapolated data from panel
 (c) assuming that $\rho_{xx}$
   is linearly dependent on $B$ in  (c) for $H||$($ab$) (open squares).
 The dashed line is a linear fit to the data. The arrow indicates
 the crossover to a Fermi liquid $T^2$ behaviour below $T_{\rm FL}$ (dotted line).
 (e) Quantum oscillations, $\Delta\rho_{xy}/\rho_{xy}$, of the $t$ = 58 nm flake
and two bulk single crystals in $\rho_{xy}$ for S1 and $\rho_{xx}$ for S2.
 (f) The corresponding FFT spectra from (e) for the flake and the bulk sample.
 The amplitude of the oscillations of the $t$ = 58 nm flake is multiplied by a factor of 10 in (e) and (f).
(g) The temperature dependence of the oscillation amplitude for the $t$ = 58 nm flake.
A Lifshitz-Kosevich (LK) fit (solid line) yields an effective carrier mass of $m_{\rm eff}=3.1(2)$m$_{\rm e}$
 for the hole pocket $\delta$, located at the centre of the Brillouin zone.
(h) The corresponding Dingle plot for the estimation of the slope which gives
$T_{\rm D}= 4.5(4)$~K.
 }
 \label{fig:Figure_3}
\end{figure*}

\subsection*{High-field magnetotransport}
Figs.~\ref{fig:Figure_3}(a) and (b) show the magnetotransport behaviour of a $t$ = 58 nm flake in high magnetic fields up to 37.5~T.
The Hall resistivity is observed to be strictly linear in magnetic field as expected for a perfectly compensated two-band system.
However, the longitudinal magnetoresistance exhibits an unconventional
 $B^{1.6}$ dependence for different devices (Fig.~S6 and Fig.~S7 in the SI Appendix \cite{SM}),
 similar to that found in bulk FeSe at high magnetic fields \cite{Bristow2020}.
Fig.~\ref{fig:Figure_3}(c) shows in-plane magnetotransport studies that are not affected by orbital effects,
 as the current and magnetic field are parallel to each other;
 the linear high field extrapolation is used to access the low temperature normal resistivity, as shown in Fig.~\ref{fig:Figure_3}(d).
We find that the low temperature resistivity has a linear temperature dependence to the lowest temperatures
for most of the measured flakes (Figs.~S7(f) and S8(b) in the SI Appendix \cite{SM}),
except for the one which displays quantum oscillations, as shown in  Fig.~\ref{fig:Figure_3}(d).
A crossover transition to the Fermi liquid behaviour occurs below 5~K for $t=58$~nm,
but this crossover is highly sensitive to the degree of  impurity scattering, as found in FeSe$_{1-x}$S$_x$ \cite{Bristow2020}
and Cu-substituted FeSe \cite{Zajicek2021Cu}.
Linear dependence at the lowest temperature  is
found for a  flake with $t=$100~nm (Fig.~S7(f) in the SI Appendix \cite{SM})
and it describes the resistivity behaviour below 50~K for both orthorhombic directions
in another flake
in Fig.~S8  in the SI Appendix \cite{SM}.
This behaviour is often a hallmark of scattering by spin fluctuations in the vicinity of
an antiferromagnetic critical point \cite{Rosch1999,Kasahara2010}.

At the lowest temperatures
we have detected  quantum oscillations
for one of the flakes with $t=58$~nm,
 both in $\rho_{xx}$ and $\rho_{xy}$ components
(Fig.~\ref{fig:Figure_3}(a) and (b)),
with the amplitude of the signal in the Hall component being stronger.
Fig.~\ref{fig:Figure_3}(e) shows quantum oscillations in the $\rho_{xy}$ component having
an amplitude a factor of 10 smaller than bulk single crystals \cite{Watson2015b}.
 The fast Fourier transform (FFT) spectra help to identify the extremal areas of the Fermi surface pockets normal to the applied magnetic field,
 $A_{ki}$, via the Onsager relation, $F_i = A_{ki} \hbar /(2 \pi e)$ \cite{Shoenberg1984}.
The low temperature experimental Fermi surface of FeSe is composed of one warped cylindrical hole band at $\Gamma$ with
oscillation frequencies $\beta$ ($k_z=0$, $F$ = 164 T) and  $\delta$ ($k_z=\pi/c$, $F$ = 664~T), and potentially
two warped cylindrical electron Fermi surfaces
that are located at the corners of the Brillouin zone
(see Fig.~\ref{fig:Figure_3}(f) and Fig.~\ref{fig:Figure_4}(b))
\cite{Terashima2014,Watson2015a,Coldea2017}.
Fig.~\ref{fig:Figure_3}(f) shows
 the dominant oscillation frequency of the $t=$58~nm thin flake is  595~T,
which  is likely to correspond to the largest orbit at the $Z$ point of the hole band $\delta$.
The signal from the hole bands was also found
to be dominant in the $\rho_{xy}$ for bulk crystals \cite{Watson2015b}.
The observed reduction in the size of the extremal area of the Fermi surface of the thin flake
could suggest a reduction of $k_{z}$ warping due to an increase in the degree of two-dimensionality in the thin flakes.
The cyclotron-averaged effective masses of the quasiparticles
 extracted from the temperature dependence of the amplitude of the quantum oscillations in Fig.~\ref{fig:Figure_3}(g)
 (using raw data from Fig.~S6(a) in the SI Appendix \cite{SM})
 is found to be $\sim 3.1(2)~m_e$,  slightly lighter than $\sim 4.5(1)~m_e$,
 found for  the bulk $\delta$ pocket   \cite{Terashima2014,Watson2015a}.

\subsection*{Scattering}
\begin{figure}[t!]
\centering
\includegraphics[trim={0cm 0cm 0cm 0cm},width=0.8\linewidth,clip=true]{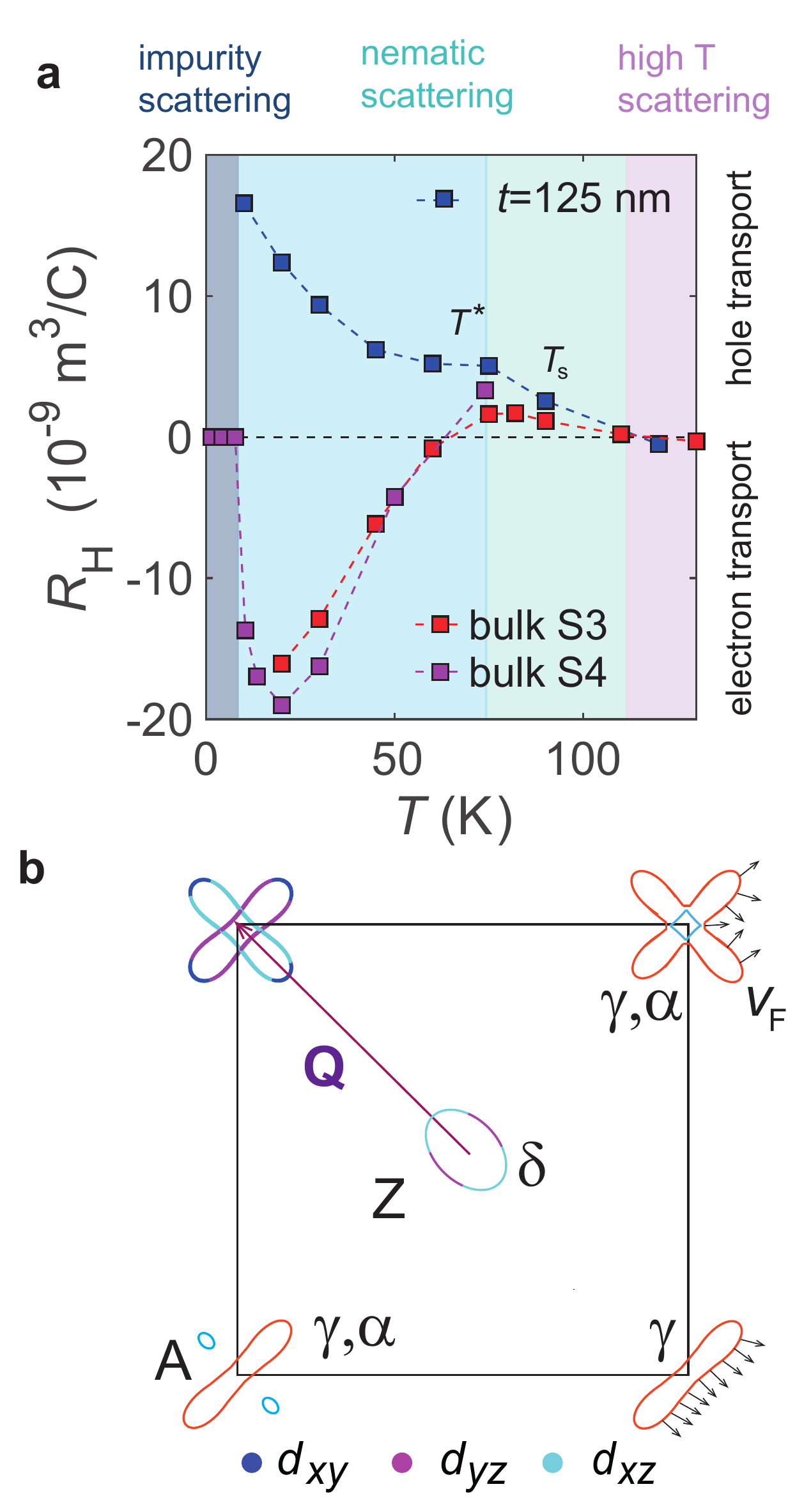}
\caption{{Anomalous transport of FeSe.}
(a) The temperature dependence of the Hall coefficient in
bulk single crystals (S3 from Ref.~\cite{Watson2015b} and S4 from \cite{Bristow2020})
as compared with a clean $t=125$~nm flake of FeSe,
with similar $RRR \sim 32$ value (Fig.~S4(c) in the SI Appendix \cite{SM}).
The potential different regimes of scattering are illustrated by different coloured regions,
  as  detailed in Ref.~\cite{Breitkreiz2014}.
Inside the nematic phase, the orbital ordering and
 spin-orbit coupling can generate sizable in-plane anisotropy in susceptibility
 below $T^*$ \cite{Li2020}).
(b) Schematic Fermi surface of FeSe at low temperatures based on quantum oscillations for $k_z=\pi/c$ \cite{Coldea2019,Coldea2021}.
The electron orbits in the corner of the Brillouin zone reflect different representations
induced by orbital ordering, spin-orbit coupling and strain \cite{Watson2017strain,Yi2019,Zhang2016,Coldea2021},
from two electron pockets (top corners) to a single peanut pocket (bottom right) or a single peanut with two small pockets (bottom left).
The small arrows indicate the variation of the Fermi velocity  when the magnetic field is normal to the
plane of the pocket that could generate electron and hole contributions to the Hall effect \cite{Ong1991}.
The dominant orbital character of the Fermi surface is shown in the top left corner.
Hot spots can be generated at the crossing point
between elliptical electron pockets \cite{Breitkreiz2014}.
}
 \label{fig:Figure_4}
\end{figure}

In order to quantify how the amplitude of quantum oscillations is affected by impurity scattering,
we estimate the Dingle temperature, $T_{\rm D}$,
as shown in Fig.~\ref{fig:Figure_3}(h)  and detailed in the SI Appendix \cite{SM} and Ref.~\cite{Carrington2011}.
The slope gives a $T_{\rm D}= 4.5(4)$~K  which corresponds to  a
 quantum mean free path of $\ell_q \sim 140 $~\AA~
 for the $t=58$~nm device.
The quantum scattering time,  $\tau_q = \hbar$/($2\pi k_{B}T_{\rm D}$),
corresponds to the time taken to fully randomise the linear momentum of an electron;
 it is found to be  $\tau_q=0.27(2)$~ps, corresponding to a quantum hole mobility of  $\mu_{q} \sim 158$ cm$^2$/Vs.
This quantum scattering time for the $\delta$ hole pocket is almost a factor of 3 shorter than
in bulk FeSe,  where $\tau_q= 0.7(1)$~ps,
as shown in Fig.~S2 in the SI Appendix \cite{SM}.
This could indicate an increase in the impurity and surface
scattering in this thin flake, as its $RRR$ is smaller than the bulk
(Fig.~\ref{fig:Figure_4}(b)).
Furthermore, the classical mobility
from the two-band model  yields a value close to $\mu_h=$1017~cm$^{2}$V$^{-1}$ s$^{-1}$,
 corresponding to a classical scattering time of $\tau_h \sim $1.8~ps, using the effective mass from quantum oscillations of 3.1(2)~m$_e$.
The large difference (a factor of 6) between the two scattering
occurs due to the sensitivity of the quantum mobility to both small and large angle scattering events,
while the transport mobility is mainly dominated by large angle scattering \cite{Narayanan2015}.

\section*{Discussion}
The unconventional behaviour in magnetotransport  of  thin flakes of FeSe
suggests the strong sensitivity of the quasiparticle scattering and
Fermi surface inside the nematic phase (Fig.~\ref{fig:Figure_4}(a)).
Below $T_{\rm s}$, there are significant changes in the shape of the Fermi
surface caused by  orbitally-band dependent shifts \cite{Coldea2017,Coldea2021}.
The 3D hole pocket, centered at the Z point, is pushed
below the Fermi level due to orbital ordering,
and the in-plane pockets become strongly elongated (Fig.~\ref{fig:Figure_4}(a)).
This change of the in-plane anisotropy is not expected to change the carrier density, $n$,
by a factor of 2 inside the nematic phase \cite{Coldea2017,Coldea2021},
which normally occurs when the Fermi surface undergoes a significant reconstruction,
as for BaFe$_2$As$_2$ in the presence of the SDW phase \cite{Terashima2011}.
Thus,  the drastic changes in the Hall coefficient and carrier densities of FeSe,
imply either that certain charge carriers inside the nematic phase scatter much more,
or get redistributed onto more localized $d_{xy}$ bands, and thus do not participate in conduction.
Additionally,  the small quasi-two dimensional Fermi surface pockets could also suffer topological
changes inside the nematic phase (Fig.~\ref{fig:Figure_4}b), as found in thin films of FeSe \cite{Zhang2016}
or induced by small applied strain \cite{Watson2017strain}.

The Hall coefficient, in the low field limit,  is very sensitive to the momentum-dependent scattering and the curvature of the Fermi surface.
The Fermi surface topology and the changes in curvature from convex to concave around
a pocket, leads to the variation of the scattering path vector, ${\bf l_{\bf k}}={\bf v_{F}} \tau_{\bf k}$.
The enclosed area swept by the ${\bf l_{\bf k}}$ vector, as {\bf k} moves around the Fermi surface,
 will change sign, which directly affects the sign of the Hall conductivity \cite{Ong1991}
 (see Fig.~S10 in the SI Appendix \cite{SM}).
The Fermi surface of FeSe at low temperatures has an elliptical hole pocket
and one or two electron pockets, as represented in Fig.~\ref{fig:Figure_4}(b).
The difference between the Hall coefficient of bulk and thin flakes in Fig.~\ref{fig:Figure_4}(a)
could be linked to a topological change of the Fermi surface, induced by orbital ordering \cite{Coldea2021};
this could transform a flower-shape pocket, with convex and concave curvatures and negative Hall coefficient,
into an elongated ellipse in thin flakes, with a convex curvature that gives a positive Hall coefficient
  (Fig.~\ref{fig:Figure_4}(b) and Fig.~S10 in the SI Appendix \cite{SM}).
The small electron pocket $e_2$ pocket of the bulk ($\alpha$ pocket)
is absent in the mobility spectrum of thin flakes of FeSe.
As the inner electron band is located very close to the Fermi level, it is highly sensitive
to small changes in energy ($\sim 3$meV),  upon reducing the thickness of the flakes, and under applied uniaxial strain.
 This pocket already disappears in thin films of FeSe \cite{Zhang2016} (Fig.~\ref{fig:Figure_4}(b)),
due to the orbital and momentum-dependent energy splitting at the M point
that is larger for thin films ($\sim 70$~meV)  \cite{Zhang2016} than in bulk ($\sim 50$~meV) \cite{Coldea2017}.
 Furthermore, ARPES studies of bulk FeSe under strain,
 that probe the surface and  bulk layers up to $\sim 10$~\AA~ \cite{Damascelli2003},
 usually detects a single electron pocket in the corner of the Brillouin zone that can be induced by applied strain
 \cite{Watson2017strain,Yi2019} (Fig.~\ref{fig:Figure_4}(b)).

The strong disparity between the hole and electron mobilities behaviour (up to a factor of 6 at 10~K),
and the observation of the rather temperature-independent mobility of the electron carriers below 50~K
could imply an enhancement of the orbitally-averaged effective masses and/or orbitally-dependent scattering that
affect mainly the electron pockets.
 The proximity to a van-Hove singularity caused
by orbitally-dependent shifts in FeSe can also
 amplify the small angle scattering processes
  for the electron pockets.
Using the low temperature mobility values of electrons
 of $\mu_e$= 231 cm$^2$/(Vs) and $\tau \sim 1.8$~ps,
from the two-band model  (Fig.~\ref{fig:Figure_3}(a) and (b))
one can estimate the effective mass to be  $\sim 13~m_{\rm e}$ for the $t=58$~nm flake.
This value is much larger  than the orbitally-averaged effective mass of $\sim 7$ m$_e$
of the electron pocket $\gamma$ for bulk crystals
 (Fig.~\ref{fig:Figure_4}(b))\cite{Watson2015a}
 and it  would be difficult to detect experimentally
in quantum oscillations (Fig.~\ref{fig:Figure_3}(f)).
Assuming that the effective masses of heavy electrons
are the same for the bulk and thin flakes,
then the changes in mobilities between electron and hole carriers
could reflect an anisotropy in scattering
(classical scattering time being $\sim 0.9$~ps for electrons and 1.8~ps for holes).
Interestingly, the effective mass extracted from ARPES studies is
much smaller, $\sim 1m_{\rm e}$ for a single momentum direction \cite{Watson2015a,Coldea2021}
and it can be enhanced in thin films and flakes of FeSe up to
$\sim  4 m_{\rm e}$ both using  K-dosing of FeSe \cite{Wen2016} and ionic liquid gating  \cite{Zhang2021}.
Orbitally-dependent band shifts and renormalizations
 were detected previously in FeSe,
with the dominant $d_{xy}$ hole band being the most renormalized by a factor of $8$
as compared with a factor 2.5-3.5 for the $d_{xz}$  and $d_{yz}$ orbitals \cite{Watson2015a,Yin2011}.
Interestingly,  the part of the electron pockets with $d_{xy}$ character can
become completely incoherent  (Fig.~\ref{fig:Figure_4}(b))
and hard to detect in surface-sensitive experiments \cite{Kreisel2020}.
Alternatively, there is an exchange of $d_{xy}$ spectral weight from one electron pockets towards the other electron
pocket,  as  suggested  by recent ARPES studies under strain \cite{Cai2020}.
Furthermore, the orbitally-dependent  pairing  between electron and hole pockets
was detected from surface studies using STM measurements of FeSe \cite{Sprau2016,Kreisel2020}.

The Hall coefficient $R_{\rm H}$ of FeSe increases with decreasing
temperature and has an inflection point at $T^*\sim 75$~K,
 for both bulk and thin flakes, as shown in Fig.~\ref{fig:Figure_4}(a).
 On cooling $R_{\rm H}$ starts deviating
 and $n$ is significantly reduced inside the nematic phase,
 as if  there is a loss of  available charge carriers.
 (Fig.~\ref{fig:Figure_2}(d)).
 At the temperature $T^*$, applied strain has the weakest effect on resistivity
 and the transport anisotropy changes sign \cite{Ghini2021}.
 The change in resistivity anisotropy coincides also
with the temperature at which a large anisotropy develops in
the local spin susceptibility \cite{Liu2010}. Thus, the anisotropy of
the local magnetism affects the quasiparticle scattering and the
coherent coupling between local spins and itinerant electrons.
 The sign of the Hall coefficient is always positive in thin flakes but
 negative in the bulk below $\sim 60$~K, despite having similar $RRR$ values.
 The Hall coefficient is also positive in Cu-substituted FeSe with large
impurity scattering \cite{Gong2021,Zajicek2021Cu} and in thin films of FeSe$_{1-x}$S$_x$ \cite{Shikama2019}.
Furthermore, it also becomes positive by using
 the isoelectronic sulphur substitution in single crystals of FeSe$_{1-x}$S$_x$ (for $x>0.11$)
  \cite{Bristow2020,Huang2020,Culo2021},
due to subtle changes in the band structure and
spin-fluctuation scattering \cite{Coldea2021,Bristow2020}.
At lowest temperature, there is a crossover from inelastic to impurity-dominated
scattering, when the $n$ value normally recovers to the high-temperature tetragonal case
 (Fig.~\ref{fig:Figure_4}(a)).
  The very large increase in $R_{\rm H }$ observed
  in the thinnest $t$ = 14~nm flake
 resembles the  behaviour of thin films of FeSe
 with lower $RRR$ values \cite{Zhang2021}.
 Recently, it was shown theoretically that
the impurity scattering in FeSe can give rise to anisotropic scattering and anisotropy in resistivity  \cite{Marciani2022}.
Furthermore, a strong  role of orbital-differentiation on the temperature
dependence of $R_{\rm H}$ has also been
found in other systems, like Sr$_2$RuO$_4$  \cite{Zingl2019} and FeCrAs \cite{Lau2021}.

Despite the lack of long-range magnetic order in FeSe,
there is a large energy range of magnetic fluctuations due to the
 relatively small spin-fluctuation bandwidth
together with the low-carrier density  \cite{Wang2016}.
In zero-magnetic field in thin flakes, we detect a linear resistivity below 50~K in most flakes,
(Fig.~S8(b) in the SI Appendix \cite{SM})
 except in the cleanest samples in which a crossover to
Fermi liquid behaviour occurs, Fig.~\ref{fig:Figure_3}(d).
The linear resistivity occurs for both orthorhombic directions  consistent with
scattering by critical antiferromagnetic fluctuations in the presence of disorder
 which is strongly enhanced at {\it hot spots} on the Fermi surface where the nesting is perfect
(Fig.~\ref{fig:Figure_4}(b)) \cite{Rosch1999,Koshelev2016,Breitkreiz2013}.
In the presence of spin fluctuations,
the quasiparticle currents dressed by vertex corrections acquire the character
of the majority carriers and lead to a larger absolute Hall coefficient with a marked
temperature dependence \cite{Fanfarillo2012}.
Additionally, the localization of electrons could be enhanced by spin-fluctuations that
affect interband scattering between elliptical electron pockets, like the  N\`{e}el-type fluctuations,
as compared with interband stripe order fluctuations between holes and electrons (Fig.~\ref{fig:Figure_4}(b)).
Short-range, weak N\`{e}el fluctuations strongly suppress the
$s_{\pm}$ superconducting state and can lead to a low-$T_{\rm c}$  $d$-wave state
\cite{Fernandes2013}.

\section*{Concluding Remarks}
In summary, we have performed a detailed study of electronic transport of high quality FeSe thin flakes
 and identify an unusual localization effect
 of  negative charge carriers  inside the nematic phase.
 This disparity  between hole and electrons emphasizes
 the anomalous transport inside the nematic phase,
 driven by the subtle interplay between the changes
 in the electronic structure of a multi-band system and the unusual
 scattering processes  induced by  orbital-dependent enhanced correlations and/or anisotropic spin fluctuations.
The two-dimensional confinement of thin flakes affects the mobility of the electron-like carriers significantly
but also plays a role in their superconductivity which is suppressed.
These effects emphasize the complexity and sensitivity  of the electron pockets in FeSe-based systems
which are involved in the stabilization of a two-dimensional high-$T_{\rm c}$ superconductivity
  via electron doping  induced by interfacial effects or dosing.

\vspace{0.5cm}

{\bf Materials and Methods.}
Thin FeSe flakes were mechanically exfoliated from high quality single crystals onto silicone elastomer polydimethylsiloxane (PDMS) stamps.
Flakes of suitable geometry and thickness were then transferred onto Si/SiO$_{2}$ (300 nm oxide) substrates with pre-patterned Au-contacts using 
a dry transfer set-up housed in a nitrogen glovebox with an oxygen and moisture content $<$1 ppm.
To minimise environmental exposure, a capping layer of thin ($\sim$20 nm) hexagonal boron nitride (h-BN) was transferred on top of the FeSe flake.
 The thickness of each sample was measured using an atomic force microscope (AFM) after all magnetotransport measurements had been performed.
Magnetotransport measurements at temperatures down to 2 K and magnetic fields up to 16~T were
performed using a Quantum Design Physical Property Measurement system (PPMS) in Oxford, with high field
 measurements performed at the High Field Magnet Laboratory in Nijmegen (up to 37.5 T) in a Helium-3 cryostat.
The magnetoresistance and Hall resistivity contributions were separated by symmetrizing and antisymmetrizing the data obtained in positive and negative magnetic fields.
 The non-ideal flake and contact geometries were accounted for by numerically evaluating the resistance to resistivity conversion factors.
 Details of these calculations are provided in the Supplementary Information
 (Fig.~S1 in the SI Appendix \cite{SM}).

\vspace{0.5cm}

{\bf Acknowledgments.}
We thank Steve Simon and Siddharth Parameswar for useful discussions 
and Roemer Hinlopen for the development of the software used to estimate 
the scattering-path length for an arbitrary Fermi surface. The research was funded 
by the Oxford Centre for Applied Superconductivity at Oxford University. 
We also acknowledge financial support from the John Fell Fund of Oxford University. 
This work was partly supported by Engineering and Physical Sciences Research Council (EPSRC) 
Grants EP/I004475/1 and EP/I017836/1. L.S.F. was supported by the Bath/Bristol Centre for Doctoral 
Training in Condensed Matter Physics, under the EPSRC Grant EP/L015544. Part of this work was supported 
by High Field Magnet Laboratory– Radboud University Nijmegen/Foundation for Fundamental Research on Matter, 
members of the European Magnetic Field Laboratory (EMFL), and EPSRC via its membership to EMFL Grant EP/N01085X/1. 
A.A.H. acknowledges financial support of Oxford Quantum Materials Platform Grant EP/M020517/1. 
Z.Z. acknowledges financial support from EPSRC Studentships EP/N509711/1 and EP/R513295/1. 
A.I.C. acknowledges EPSRC Career Acceleration Fellowship EP/I004475/1.

\vspace{0.5cm}

{\bf Data availability}
The data that support the findings of this study are available through the open access 
data archive at the University of Oxford (ORA) (https://doi.org/10.5287/bodleian:X5GgyEj1O)
Additional information about the data can be addressed  to the corresponding author.

\newpage
\clearpage
\setcounter{equation}{0}
\setcounter{figure}{0}
\setcounter{table}{0}
\makeatletter
\renewcommand{\theequation}{S\arabic{equation}}
\renewcommand{\thefigure}{S\arabic{figure}}

\widetext
\begin{center}
\textbf{\large Supplemental Information: Unconventional localization of electrons inside a nematic electronic phase} 
\end{center}

\begin{center}
{\bf Geometrical corrections}
\end{center}
\noindent
The exfoliation of rectangular-shaped flakes onto pre-patterned contacts leads to samples with rather non-ideal geometries.
The current flow is inevitably quite inhomogeneous and the contacts extend a long way underneath the flake, violating the assumptions of the van der Pauw approach commonly used for irregularly-shaped samples. As a consequence the longitudinal and Hall resistivities have been estimated from finite difference solutions of the transport equation for a realistic sample and contact geometry.

For our two-dimensional problem the equation to be solved is given by:
\begin{equation}
\vec{E}+\rho_{xy} \vec{J} \times \hat{z}=\sigma_{xx}^{-1} \vec{J},
\label{eqn:S1}
\end{equation}
\noindent
where $\vec{E}$ is the 2D electric field and $\vec{J}$ the 2D current density, $\hat{z}$ is unit vector perpendicular to the sample
and is the direction of the applied magnetic field, $\vec{B}$=$\mu_0 \vec{H}$;
$\sigma_{xx}$ and $\rho_{xy}=(B)/(n_{2D} e)$ are the sheet conductivity and Hall resistivity of the flake respectively,
with $n_{2D}$ the carrier concentration.

\begin{figure*}[htbp]
 \centering
  	\includegraphics[trim={0cm 0cm 0cm 0cm}, width=1\linewidth,clip=true]{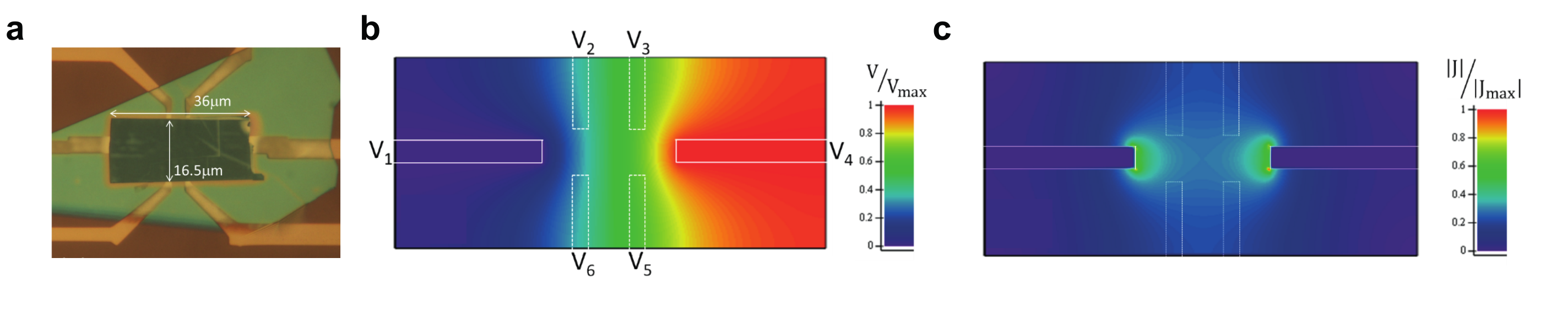}
 \caption{{\bf Simulation of the current and voltage distribution in a thin flake device.} Optical image of a FeSe flake device with $t=125$~nm.
 (b) Calculated potential distribution, $V$,
  and (c) the modulus of the calculated current density, $J$, for a $ 144 \times 66$ site model of the sample geometry
  shown in (a) with $g$ = $\sigma_{xx} \rho_{xy}$ = 0.01.
  Superimposed solid and dotted lines indicate the boundaries of the underlying current and voltage leads, respectively.}
 \label{SM:Rxx_Rxy_device}
\end{figure*}

 Fig.~\ref{SM:Rxx_Rxy_device} illustrates an example used to solve Equation~\ref{eqn:S1} in terms of the potential, $V$, on a $ 144 \times 66$ site grid.
 We impose the following boundary conditions along the left and right edges of the flakes
 where $J_{x}=0$, the two components of the electric field satisfy $E_{x} = -(\sigma_{xx} \rho_{xy} ) E_{y}$, whereas
 along the top and bottom edges where $J_{y} = 0$, the two components of the electric field satisfy $E_{y}=(\sigma_{xx} \rho_{xy} ) E_{x}$.
 Once the solution has converged the current passing through the device, $I$, is calculated by summing the following
 expression for $J_{x}$ down a vertical line through the centre of the device.
\begin{equation}
J_{x}=\frac{\sigma\left(E_{x}+\left(\sigma_{xx} \rho_{xy}\right) E_{y}\right)}{\left(1+\left(\sigma_{xx} \rho_{xy} \right)^{2}\right)}
\end{equation}

We assume that the current leads are strongly coupled to the sample and arbitrarily fix $V_{1} = 0V$ and $V_{4} = 1V$ everywhere in the flake directly above them.
In contrast we assume that the voltage contacts are only weakly coupled to the flake reflecting the strong anisotropy of our layered materials
(a factor larger than 4 which increases with the reduction in the flake thickness  \cite{Farrar2020}),
and calculate the potential at each of the voltage leads as the average of the potential in the flake directly above it.
It is then straightforward to show that the scaling factors (which are typically in the range 1-3) that must be applied to obtain the longitudinal resistivity, $\rho_{xx} = 1 / \sigma_{xx}$, and the Hall resistivity, $\rho_{xy}$, from the experimentally-measured longitudinal resistance, $R_{xx}$, and Hall resistance, $R_{xy}$, are:
\begin{equation}
\begin{array}{l}
\rho_{xx}=\left(\frac{I}{{V}_{3}-{V}_{2}}\right) \times R_{xx}, \\
\rho_{xy}=\left(\frac{I \cdot g}{{V}_{2}-{V}_{6}}\right) \times R_{xy}
\end{array}
\label{eqn:S4}
\end{equation}
\noindent
Here $g = \sigma_{xx} \rho_{xy}=$0.01 is a parameter that is used in the numerical calculation to effectively define
the magnetic field used to solve for the Hall voltage.
Using the definitions in Equations~\ref{eqn:S4}  then the conversion factors for
each resistivity component are $G_{xx}=I/(V_3-V_2)$
 and $G_{xy}=I \cdot g/(V_2-V_6)$.
Since Equation~\ref{eqn:S1} yields a Hall voltage that is linear
in magnetic field, $H$, the scaling factor for the Hall resistivity calculated
in Equation~\ref{eqn:S4} does not depend on this.

\newpage

\begin{center}
\textbf{Two-band model}
\end{center}
\noindent
Considering a multi-band system in which current is applied along the $x$  in-plane axis of a sample with a magnetic field applied along the $z$ out-of-plane axis, the total resistivity is given by $\rho=\left(\sum_{i} \rho_{i}^{-1}\right)^{-1}$ assuming parallel network resistor which in an applied magnetic field $B$ leads to:
\begin{equation}
\rho_{i}=\left(\begin{array}{cc}
\frac{1}{\sigma_{i}} & -R_{i} B \\
R_{i} B & \frac{1}{\sigma_{i}}
\end{array}\right)
\end{equation}
where the conductivity $\sigma_{i}=\left|n_{i} e \mu_{i}\right|$ and the Hall coefficient $R_{i}=-1 / n_{i} e$ contain the carrier densities $n_{i}$ and mobilities $\mu_{i}$ of $i$ number of bands.
\begin{equation}
\rho_{i}=\frac{1}{\left|n_{i} e \mu_{i}\right|} \left(\begin{array}{cc}
1 & -\mu_{i} B \\
\mu_{i} B & 1
\end{array}\right)
\end{equation}
\noindent
For a two-carrier system, the different components of the resistivity tensor are given by:
\begin{equation}
\rho_{x x}=\frac{\left(\sigma_{1}+\sigma_{2}\right)+\sigma_{1} \sigma_{2}\left(\sigma_{1} R_{1}^{2}+\sigma_{2} R_{2}^{2}\right) B^{2}}{\left(\sigma_{1}+\sigma_{2}\right)^{2}+\sigma_{2}^{2} \sigma_{2}^{2}\left(R_{1}+R_{2}\right)^{2} B^{2}}
\end{equation}
\noindent
and
\begin{equation}
\rho_{x y}=B\frac{\left(\sigma_{1}^{2} R_{1}+\sigma_{2}^{2} R_{2}\right)+\sigma_{1}^{2} \sigma_{2}^{2} R_{1} R_{2}\left(R_{1}+R_{2}\right) B^{2}}{\left(\sigma_{1}+\sigma_{2}\right)^{2}+\sigma_{1}^{2} \sigma_{2}^{2}\left(R_{1}+R_{2}\right)^{2} B^{2}}
\end{equation}
\noindent
FeSe in the tetragonal phase can be described as a compensated two-band system,
and we assign the conductivities  to correspond to a hole $\sigma_h$ and electron band $\sigma_e$, respectively.
Compensation requires that $R_{1}$ = -$R_{2}$ and $n=n_{e}=n_h$ leading to a simplified form of the expression above

\begin{equation}
\rho_{xx}=\frac{1+\frac{1}{(n e)^{2}} \sigma_{h} \sigma_{e} B^{2}}{\left(\sigma_{h}+\sigma_{e}\right)}
\end{equation}
and
\begin{equation}
\rho_{xy}=\frac{\frac{1}{(n e)}\left(\sigma_{h}-\sigma_{e}\right) B}{\left(\sigma_{h}+\sigma_{e}\right)}.
\end{equation}

By fitting simultaneously
the magnetic field dependence of both the longitudinal magnetoresistance, $\rho_{xx}$,
 and the transverse Hall component, $\rho_{xy}$, to the above equations,
 the carrier density $n$ and the two mobilities $\mu_{e}$ and $\mu_{h}$ can be extracted.

\begin{center}
\textbf{The mobility spectrum}
\end{center}
\noindent

The mobility spectrum has been developed to eliminate the need for making a priori assumptions
on the transport parameters in a multicarrier system.
The mobility spectrum analysis has been used extensively in multi-band semiconductors,
in multi-band topological semi-metals \cite{Zhao2021} and more limited in multi-band iron-based superconductors
 \cite{Huynh2014,Huynh2016}.
The conductivity of a multi-band system is given in terms of
a mobility spectrum, based on a method described in Ref.~\onlinecite{Beck1987}
This work is part of a separate report related to the mobility spectrum of  FeSe$_{1-x}$S$_x$ \cite{Humphries2016}.

\begin{equation}
\label{eq:sighat}
\hat{\sigma}(B)=\sigma_{xx}+i\sigma_{xy}=\int_{-\infty}^{\infty}d\mu\frac{s(\mu)(1+i\mu B)}{1+\mu^2B^2}
\end{equation}
where $s(\mu)=e\mu n(\mu)$ is the  mobility spectrum of zero-field conductivity and the charge carrier densities are
functions of mobility (with electrons defined to have negative $\mu$ and $n(\mu)$).
Thus the  fundamental object of interest to identify the properties of charge carriers in a material
 becomes the conductivity spectrum $s(\mu)$, or equivalently the carrier-density spectrum $n(\mu)$.
 The mobility spectrum constructed for $t$=125~nm in Figure~2(a) is expanded
 in different components in  Fig.~\ref{SM:FigureSM_mob_spectrum}. The initial parameters
 proposed by the mobility spectrum, together with compensation of the charge carriers,
are used to extract the discrete fitted parameters in Figs.~2(d),(e) and (f).
 The behaviour of the mobilities and carrier densities as a function
 of temperature are consistent between the two approaches.

The mobility spectrum could be a powerful tool
to identify the conduction in multiple band systems with different mobility which appear as distinct peaks in $s(\mu)$,
as long as the condition $\mu B < 1$ is satisfied.
Its fundamental description of the electrical transport  does
not require advance knowledge about the band structure or the scattering mechanisms.
 A distribution of relaxation times  results in a distribution of mobilities and
   broadening of the corresponding peak in $s(\mu)$.
  Therefore, the shape of the peak in $s(\mu)$ reveals
not only an average of relaxation times ( $ \langle \tau \rangle^2/(\tau)^2$),
but the actual distribution of relaxation times \cite{Beck1987}.

 Approximations are often used to extract the form of the mobility spectrum for a material having a
 finite data set of conductivity or resistivity measurements as $s(\mu)$ is never fully constrained on the infinite set of basis functions.
 Discrete approximations can often add ghost peaks in the spectrum,
 as well as the introduction of bias with a necessary predetermined range of interest for mobility. Other techniques, used
 in the case of iron-based superconductors, fit analytic forms to the data which can be directly transformed \cite{Huynh2014,Huynh2016}.
These approaches have the disadvantages of losing some fidelity of the data and require an analytic continuation of the
 fitted function to infinite magnetic field values.


\begin{center}
\textbf{The Lifshitz-Kosevich equation}
\end{center}
\noindent
The amplitude of quantum oscillations can be described by the Lifshitz-Kosevich equation \cite{Lifshitz1958,Shoenberg1984}:
\begin{equation}
\begin{array}{c}
\Omega=\left(\frac{e}{2 \pi \hbar}\right)^{3 / 2} \frac{e \hbar V B^{5 / 2}}{m^{*} \pi^{2}} \sum_{A_{\text {ext }}}\left|\frac{\partial^{2} A_{k, i}}{\partial k_{\perp}^{2}}\right|^{-1 / 2} \\
\sum_{p=1}^{\infty} p^{-5 / 2} R_{T} R_{\rm D} R_{S}, \cos \left(2 \pi p\left(\frac{F}{B}-\frac{1}{2}+\frac{\phi_{B}}{2 \pi}\right)\left[\pm \frac{\pi}{4}\right]\right),
\end{array}
\end{equation}
\noindent
where $R_{T}$, $R_{\rm D}$, and $R_{S}$ are damping terms. The first sum extends over all extremal Fermi surface
areas $A_{k,i}$ perpendicular to the applied field, while the second over all $p$ harmonics of the fundamental
oscillation frequency. These oscillations are periodic in 1/$B$ with a frequency which is determined by orbits on the Fermi Surface that enclose locally extremal momentum space area, where $F_{i} = \frac{\hbar}{2 \pi e}A_{k,i}$.

The first damping term, $R_{T}$, accounts for the thermal broadening from the Fermi-Dirac distribution with temperature and is given by:
\begin{equation}
\begin{aligned}
R_{T} &=\frac{X}{\sinh (X)} \\
X &=\frac{2 \pi^{2} k_{B} T \mathrm{pm}^{*}}{e \hbar B}
\end{aligned}
\label{eqn:thermal}
\end{equation}
\noindent
where $m^{*}$ is the quasiparticle effective mass. $R_{T}$ depends on the ratio $X \propto k_{B} T / \hbar \omega_{c}$, where $\omega_{c}=e B / m^{*}$ is the cyclotron frequency. These two equations contain the temperature dependence of the amplitude of oscillation, and by fitting data to Equation \ref{eqn:thermal} the effect mass of $m^{*}$ can be extracted.

\begin{figure*}[htbp]
	\centering
\includegraphics[width=0.6\linewidth,trim={0cm 0cm 0cm 0cm}]{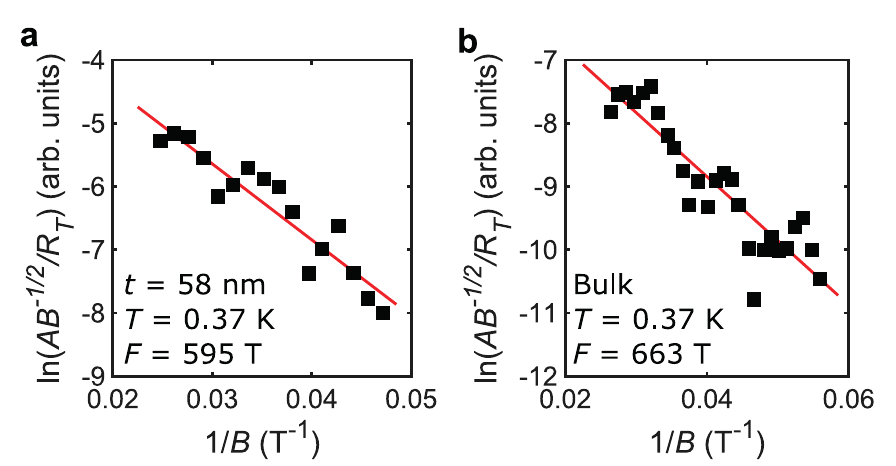}
	\caption{{\bf Estimation of the quantum scattering time from Dingle plots. }
Dingle plots to extract the Dingle temperature, $T_{\rm D}$,  from the slope and accounting
for the $R_T$  and $R_{\rm D}$ field dependence
for (a) a $t = 58$ nm device ($T_{\rm D} \sim $~4.5(4)~ K) and (b) a bulk single crystal S1 ($T_{\rm D} \sim $~1.6(2)~K).}
	\label{SM:Dingle_plot}
\end{figure*}

The second damping term, $R_{D}$, is accounts for the impurity scattering of the electrons and is given by
\begin{equation}
R_{\rm D}=\exp \left(-\frac{\pi m_{b}}{e B \tau}\right)=\exp \left(-\frac{2 \pi^{2} m_{b} k_{B} T_{\rm D}}{e \hbar B}\right),
\label{eqn:Dingle}
\end{equation}
\noindent
where $\tau$ is the scattering time, $m_{b}$ is the band mass, and $T_{\rm D}=\hbar / 2 \pi k_{B} \tau$ is the Dingle temperature.
This equations is determined by the impurity scattering rate which acts to exponentially dampen the amplitude of the quantum oscillation in magnetic field.
As the band structure calculations that provide the band mass cannot capture the Fermi surface of FeSe correctly \cite{Watson2017a},
we use $m^*=m_{b}$ to determine the Dingle temperature. The values of the bulk were taken from previous reports \cite{Watson2015a,Coldea2019}.
The third damping term, $R_{S}$, accounts for the Zeeman splitting of Landau levels.

\begin{figure*}[htbp]
	\centering
	\includegraphics[trim={0cm 0cm 0cm 0cm},width=0.8\linewidth]{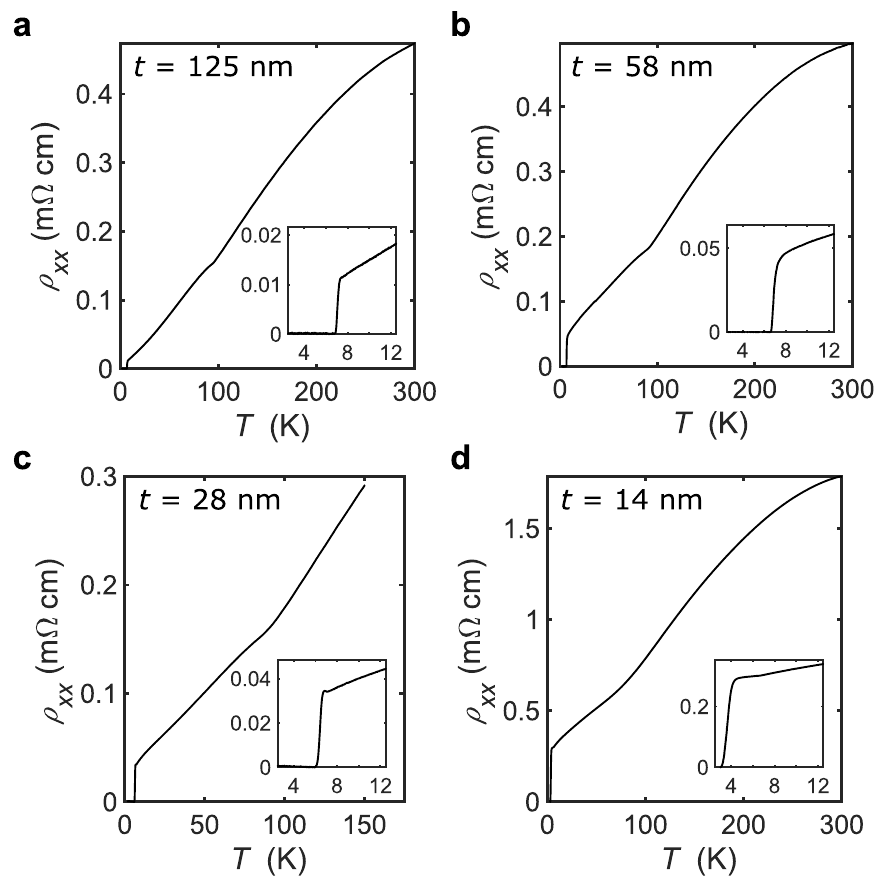}
	\caption{{\bf Resistivity in zero-magnetic field of thin flakes devices.}
Temperature dependence of the zero-field
resistivity for four thin flake devices with different thicknesses
(a) $t=$125~nm, (b) $t=58$~nm, (c) $t=28$~nm and (d) $t=$14~nm.
The insets show the low temperature superconducting transition region
and highlights that at low temperature
the resistivity has a rather linear dependence for most of the thin flakes samples.
Additional resistivity curves for other thin flakes were previously reported in  Ref.~\cite{Farrar2020}.
 A small increase in resistivity of the thinnest flakes is detected in
(c) and (d), which is often interpreted as a signature of Anderson localization due to disorder.
However, insulating behaviour was detected in the ultra-thin limit of FeSe flakes
 below 9~nm \cite{Farrar2020,Zhu2021}
 and the number of charge carriers
 remain unchanged upon reducing
 thickness (see Fig.~2(d) and Ref.~\cite{Zhu2021}).
}
	\label{SM:Rho_Ttdep}
\end{figure*}

\begin{figure}[htbp]
	\centering
	\includegraphics[trim={0cm 0cm 0cm 0cm},width=1\linewidth]{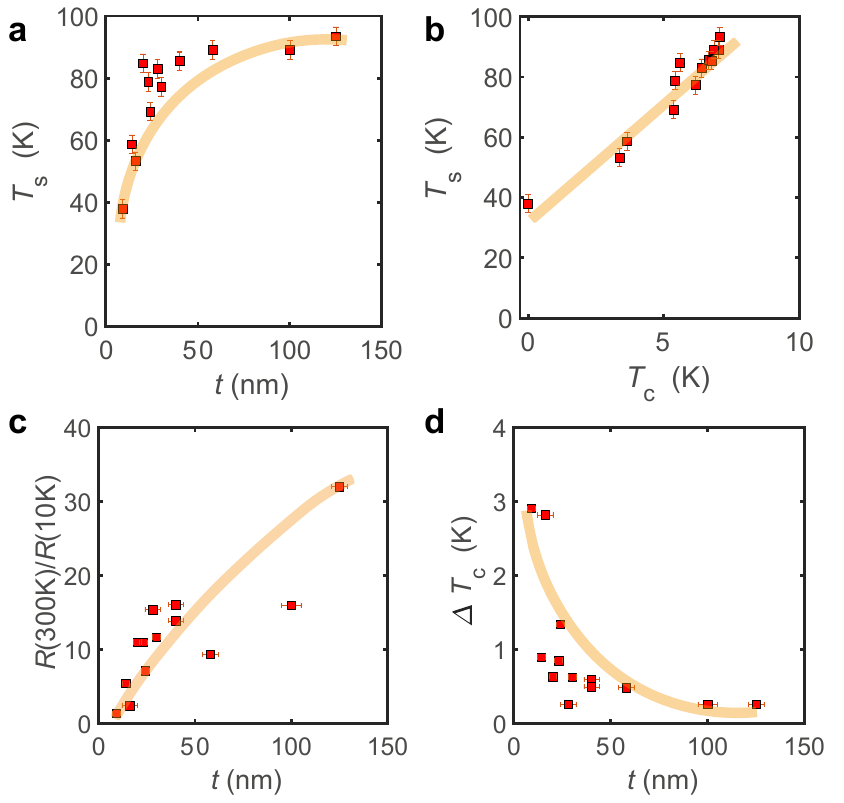}
	\caption{{\bf Transport parameters of FeSe thin flakes.}
(a) Thickness dependence of the nematic transition $T_{\rm s}$
defined as the minimum in $d\rho_{xx}/dT$.
The value of the transition of the bulk FeSe single crystals is $T_{\rm s}$=89~K.
(b) The linear relationship between $T_{\rm s}$ and
the critical temperature, $T_{\rm c}$, defined here as the middle of the superconducting transition.
The value of the critical temperature of the bulk FeSe single crystals is
always higher $T_{\rm c} \sim 8.7$~K.
(c) The thickness dependence of the residual resistivity
ratio,  defined as the ratio between the resistivity at 300~K and 10~K,
$RRR= \rho$(300~K)/$\rho$(10~K). The $R$ for the bulk
are is around 32 \cite{Bristow2020}.
 (d) The thickness dependence of the width of the superconducting
 transition, $\Delta T_{\rm c}=T_{\rm c, on}-T_{\rm c, off}$.
 The solid lines in all panels are guides to the eye.
}
	\label{FigSM:TF_parameters}
\end{figure}

\begin{figure*}[htbp]
	\centering
\includegraphics[width=\linewidth]{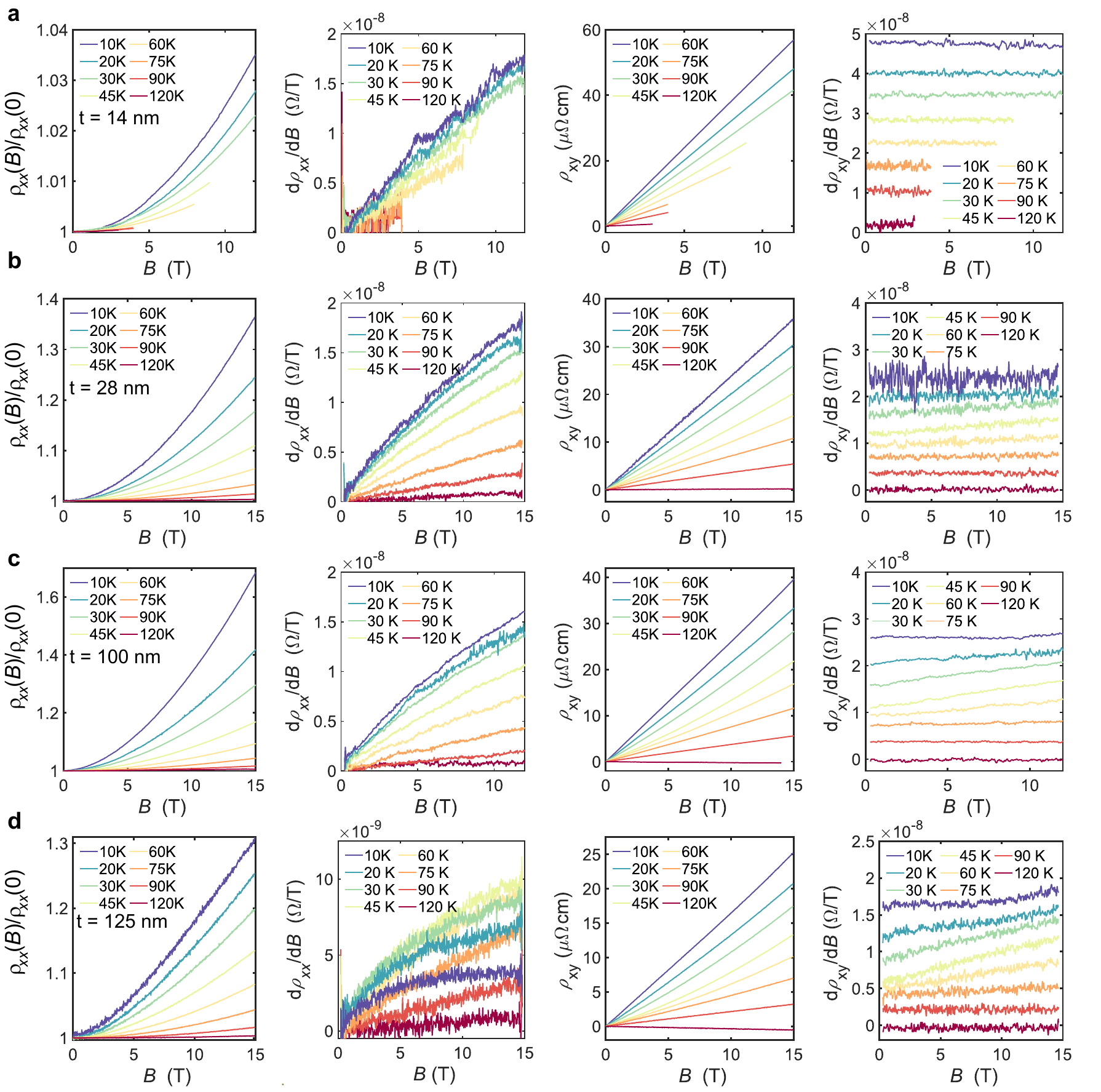}
	\caption{{\bf The magnetotransport data of different thin flake devices. } The field dependence of
the longitudinal magnetoresistance, $\rho_{xx}$,
and its derivative, $d\rho_{xx}/dB$,   the Hall component, $\rho_{xy}$,
 and its derivative $d\rho_{xy}/dB$ for different thin flake devices with thicknesses (a) $t=$14~nm,  (b)   $t=$28~nm,
 (c) $t=$100~nm and  (d) $t=$125~nm  measured at different constant temperatures.
 }
	\label{SM:Rxx_Rxy_derivatives}
\end{figure*}

\begin{figure*}[htbp]
 \centering
  	\includegraphics[trim={0cm 0cm 0cm 0cm}, width=1\linewidth,clip=true]{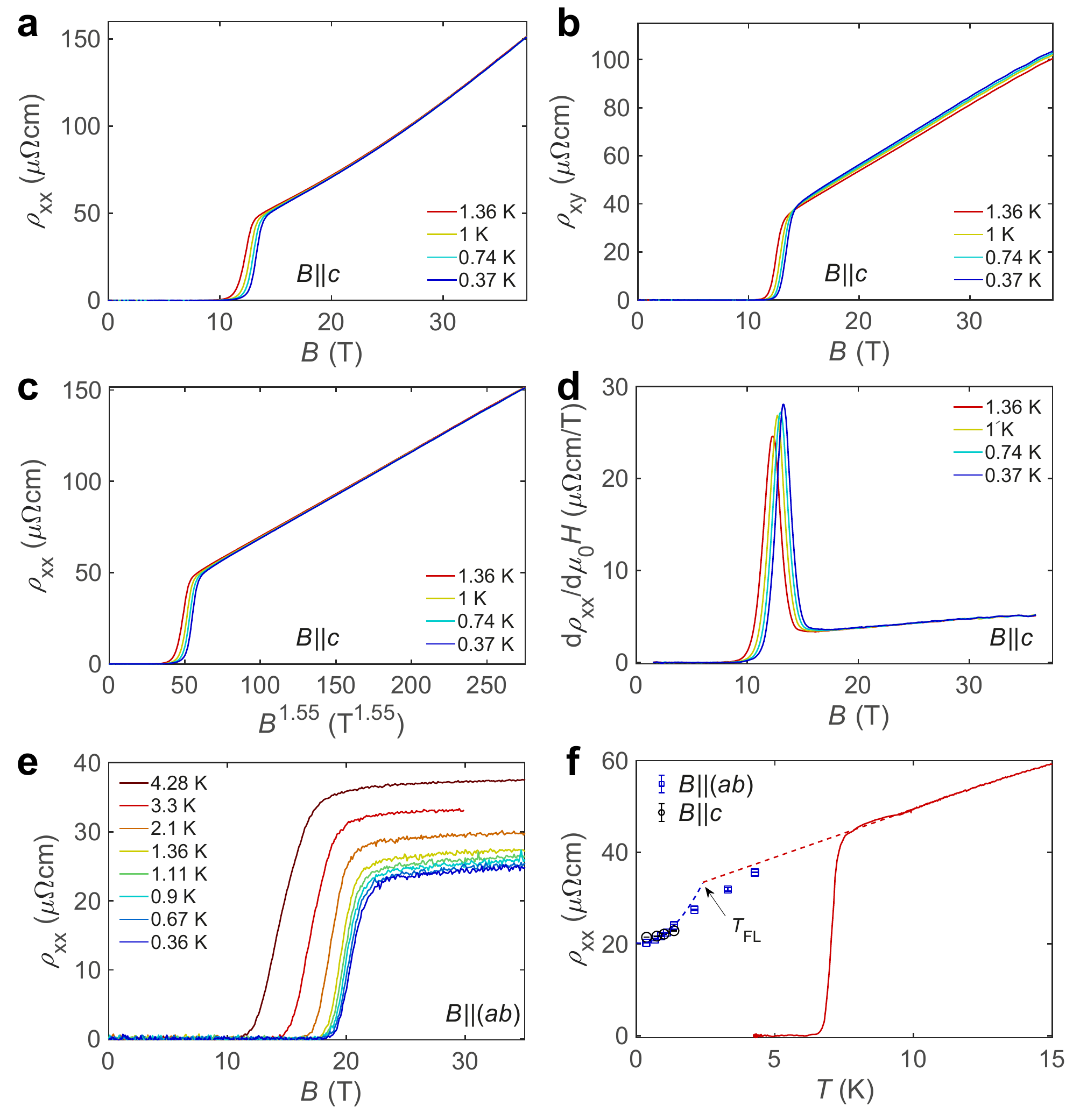}
 \caption{{\bf High magnetic field study of a $t=$58~nm device.}
  High magnetic field dependence  of (a) the longitudinal magnetoresistance, $\rho_{xx}$
 and (b) the transverse Hall component, $\rho_{xy}$, for a device with $t=$58~nm
 measured at different constant temperatures.
 (c) The longitudinal resistance versus $B^{1.6}$ shows a linear dependence, similar to  bulk single crystals \cite{Bristow2020}.
(d) The derivative of the longitudinal resistance, $d\rho_{xx}/dB$.
(e) The in-plane resistance versus magnetic field for different temperatures. (f) The low-temperature dependence
of resistance. The extrapolated values in zero field were extracted from (e) and (c).
The dashed line is a linear fit to the high temperature data that follows the extracted data to the lowest temperature.
  }
 \label{SM:Nijmegen_58nm}
\end{figure*}

\begin{figure*}[htbp]
 \centering
  	\includegraphics[trim={0cm 0cm 0cm 0cm}, width=1\linewidth,clip=true]{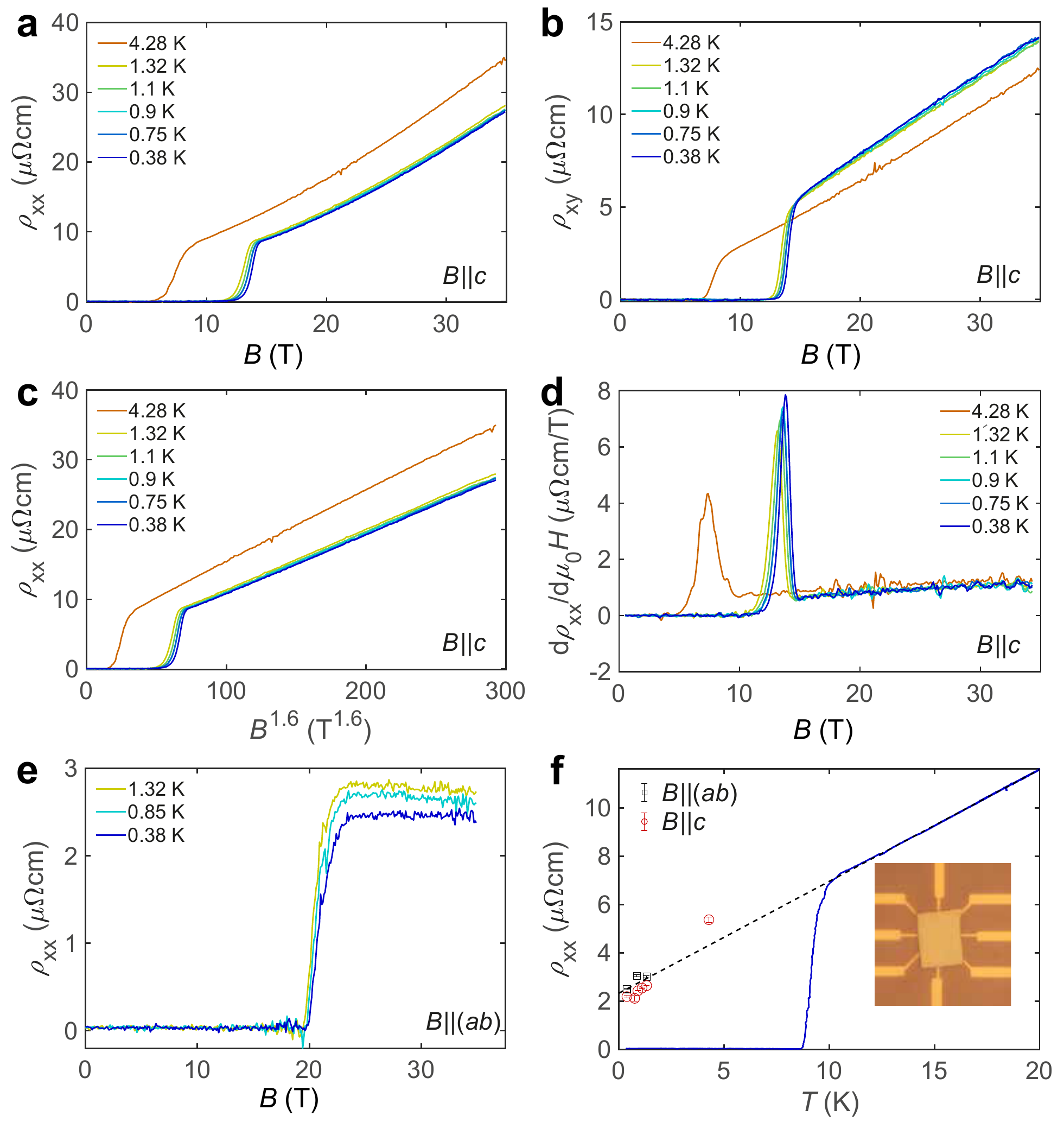}
 \caption{{\bf High magnetic field study of a $t$=100~nm device (LF8a).}
 High magnetic field dependence  of (a) the longitudinal magnetoresistance, $\rho_{xx}$
 and (b) the transverse Hall component, $\rho_{xy}$, for a device with $t$=100~nm
 measured at different constant temperatures.
 (c) The longitudinal resistance versus $B^{1.6}$ show a linear dependence, similar to bulk single crystals \cite{Bristow2020}.
(d) The derivative of the longitudinal resistance, $d\rho_{xx}/dB$.
(e) The in-plane resistance versus magnetic field for different temperatures. (f) The low-temperature dependence
of resistance. The extrapolated values in zero field were extracted from (e) and (c).
The dashed line is a linear fit to the high temperature data that follows the extracted data to the lowest temperature.
}
 \label{SM:Nijmegen_LF8a}
\end{figure*}

\begin{figure*}[htbp]
 \centering
  	\includegraphics[trim={0cm 0cm 0cm 0cm}, width=1\linewidth,clip=true]{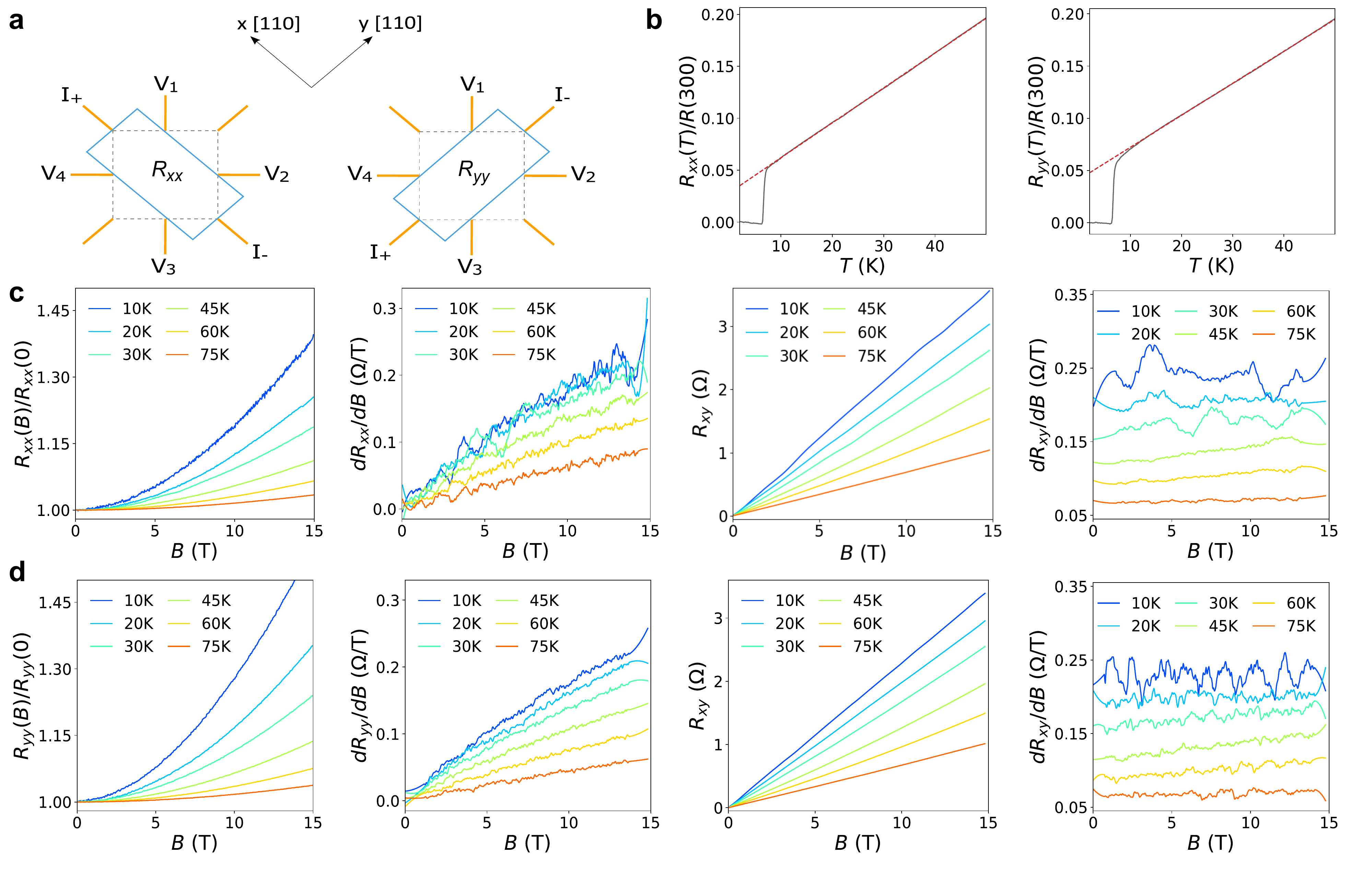}
 \caption{ {\bf Anisotropy of a multi-contact thick device $t =100$~nm.}
 (a) The different configurations
 used to probe the in-plane anisotropy.
 (b) The temperature dependence of the $R_{xx}$ and $R_{yy}$ components.
 (c) The field dependence of the longitudinal magnetoresistance, $R_{xx}$
 and the transverse Hall component, $R_{xy}$ together with their corresponding derivatives.
 (d) The longitudinal and transverse components for the second configuration
for the symmetric longitudinal component, $R_{yy}$, and the antisymmetric transverse component, $R_{xy}$
and their derivatives.}
 \label{SM:Spider_anisotropy_Oxford}
\end{figure*}

\begin{figure*}[htbp]
 \centering
  	\includegraphics[trim={0cm 0cm 0cm 0cm}, width=0.7\linewidth,clip=true]{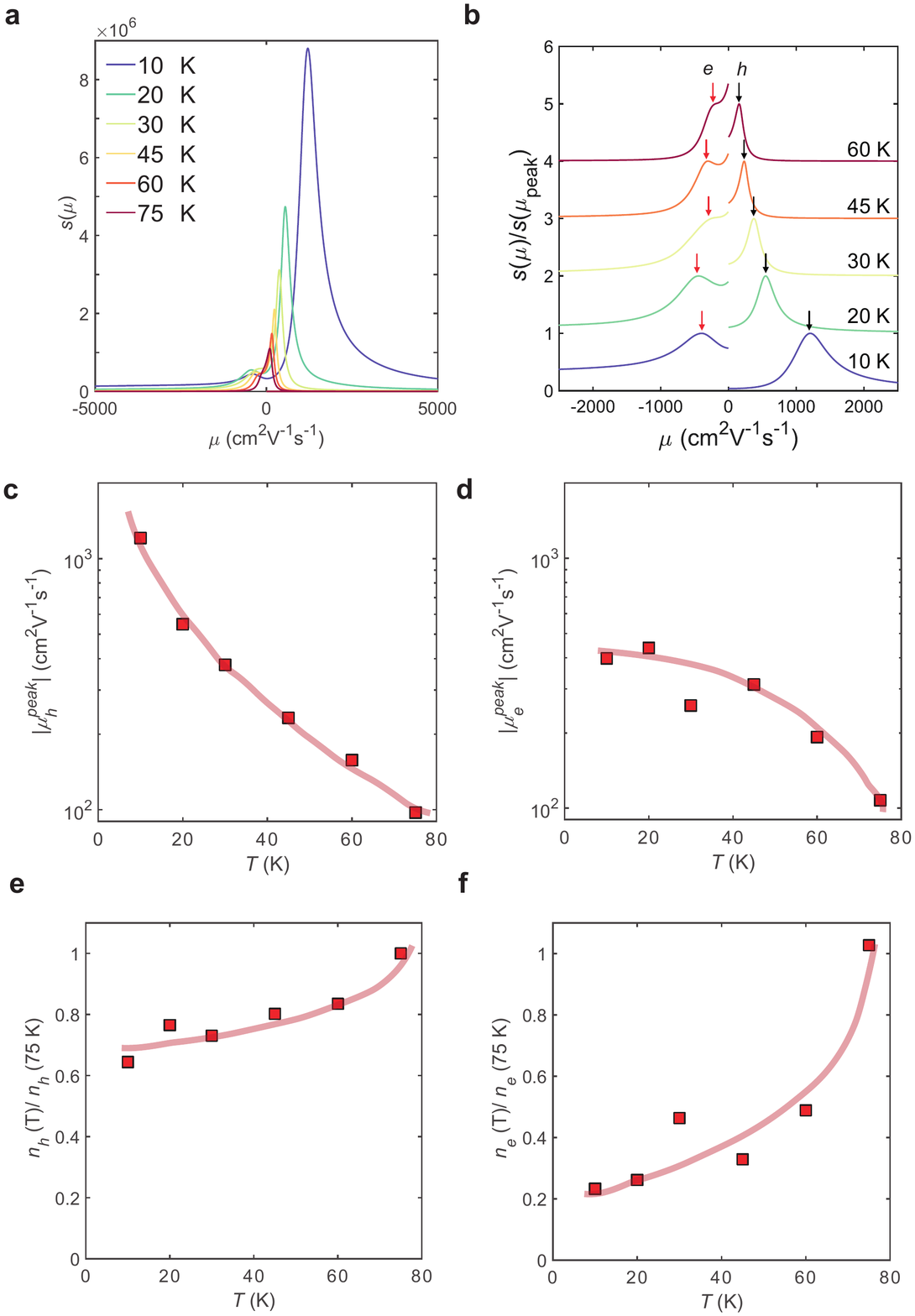}
 \caption{ {\bf The mobility spectrum of the thick device $t =125$~nm.}
 (a) The raw mobility spectrum, $s(\mu)=e \mu n(\mu )$ constructed based on the raw data
 from Fig.~1(c) and (d).
 (b) The  mobility spectrum divided to mobility value
 at the maximum peak at each temperature,  $s(\mu)/e \mu $.
 (c) and (d) The temperature dependence
 of the mobilities of positive ($\mu_h$) and negative charge ($\mu_e$) carriers extracted from the peak position from (b).
 (e) and (f) The temperature dependence
 of the number of charge carriers, $n_h$ and $n_e$  extracted from the corresponding peak positions
$n(\mu)$=$ s(\mu) / (e \cdot \mu)$ from (a) and normalized to the value at 75~K.
The solid lines are guide to the eye.
The normalized value of the carrier densities in (e) and (f)
 are affected by finite field points and other scattering
effects that cause broadening of the peaks \cite{Beck1987}.
}
 \label{SM:FigureSM_mob_spectrum}
\end{figure*}

\begin{figure*}[htbp]
 \centering
  	\includegraphics[trim={0cm 0cm 0cm 0cm}, width=0.7\linewidth,clip=true]{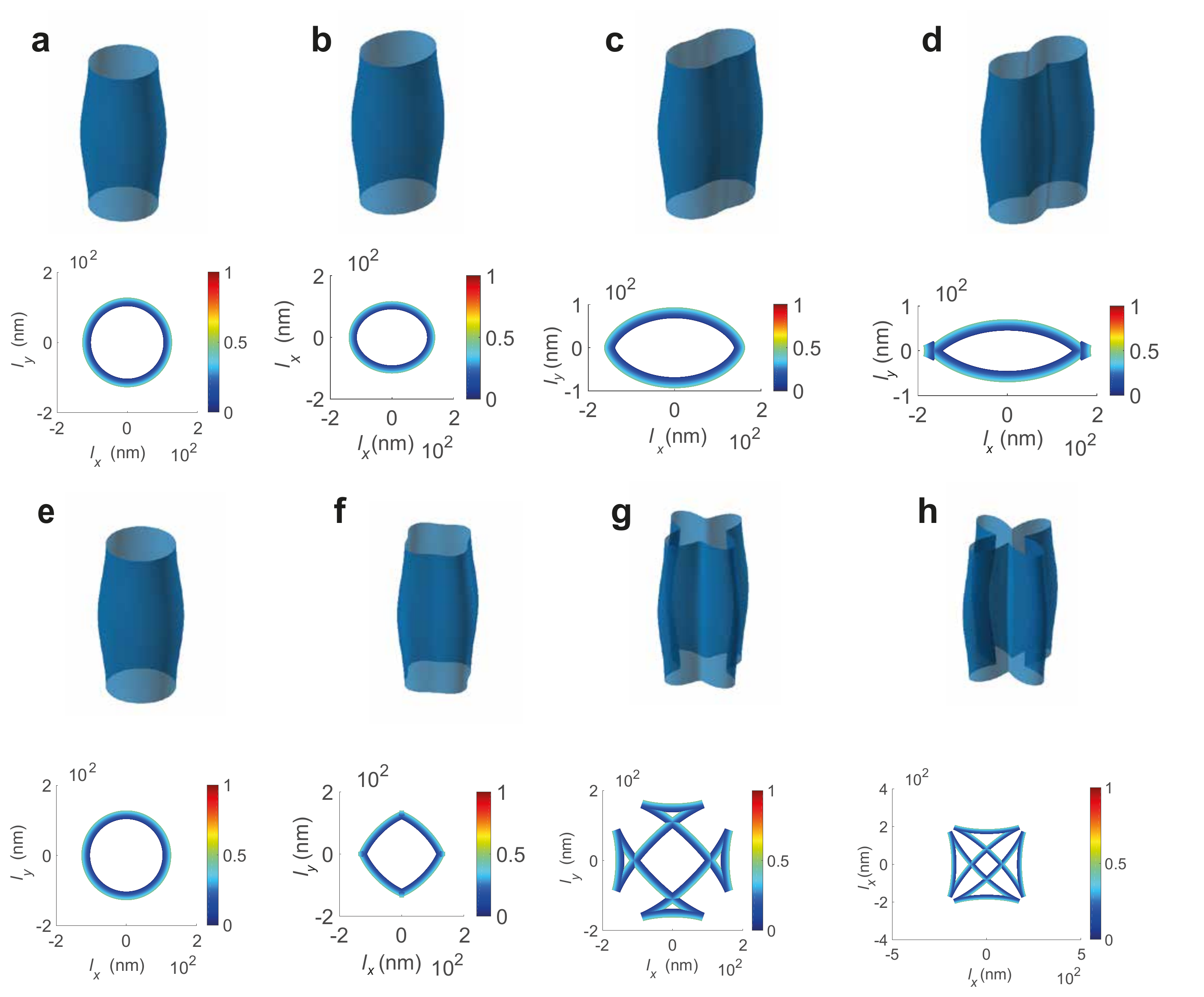}
 \caption{ {\bf Fermi surfaces and scattering path length.}
 (top panels) The evolution of the Fermi surface and the scattering length for a quasi-two dimensional Fermi surface
 expanded in cylindrical coordinates \cite{Prentice2016}  with $k_{00}=0.1$ \AA$^{-1}$,
 the interplane parameter of $k_{10}=0.01$ \AA$^{-1}$, and the in-plane two-fold symmetric
  parameter  (a) $k_{02}$=0, (b) $k_{02}$=0.01, (c) $k_{02}$=0.03, (d) $k_{02}$=0.05 \AA$^{-1}$.
 (bottom panels) The evolution of the Fermi surface and the scattering length
 of the in-plane four-fold symmetric parameter  (e) $k_{04}$=0, (f) $k_{04}$=0.01, (g) $k_{04}$=0.03, (h) $k_{04}$=0.05 \AA$^{-1}$.
The Hall conductivity is linked to the $2 \times A_{\ell}$, area swept by the scattering path length, $\ell_{\bf k}$.
In the  cases (g) and (h) the Hall conductivity can change sign as compared with the other cases.
  The scattering time is assumed isotropic,  $\tau=1$~ps \cite{Ong1991}.}
 \label{SM:FigureSM_Ong_construction}
\end{figure*}

\clearpage
\newpage

\bibliography{FeSe_bib_flakes_june2022}

\end{document}